\documentclass[conference]{IEEEtran}
\renewcommand\IEEEkeywordsname{Index Terms}
\usepackage{graphicx}
\usepackage[position=b]{subcaption}
\usepackage{tikz,xparse}
\usetikzlibrary{dsp,chains}
\usetikzlibrary{matrix}
\usepackage{mathptmx}
\usepackage{verbatim}
\usepackage{calc}% http://ctan.org/pkg/calc
\usepackage{ifthen}
\usepackage{xifthen}
\usepackage{cancel}
\usepackage{bm}
\usepackage{verbatim}
\usepackage{multirow}
\usepackage{cite}
\usepackage{relsize}
\usepackage{tikzscale}
\usepackage{color}
\usetikzlibrary{plotmarks}

\usepackage[nolist]{acronym} 
\usepackage{pgfplots}
\usetikzlibrary{arrows,shapes,graphs,graphs.standard,quotes,arrows.meta, fit}
\pgfplotsset{compat=newest}
\usetikzlibrary{matrix}

\usepackage[hyphens]{url}

\usepackage[bookmarks=false]{hyperref}
\usepackage{units}
\usepackage{amsmath, amsbsy, amssymb, latexsym }
\usepackage{comment}
\usepackage[utf8]{inputenc}
\usepackage{xcolor}
\usepackage{enumitem}
\usepackage{algorithm}
\usepackage{algpseudocode}

\usepackage{etoolbox} %To reduce vspace between authors box and text

\usepackage{flushend}

\renewcommand{\Comment}[2][.75\linewidth]{%
\leavevmode\hfill\makebox[#1][l]{$\triangleright$~#2}}
\tikzset{>=latex}



\tikzset{ vndsSplit/.style={ shape=circle split, circle split part fill={black,red}, draw, inner sep=0pt,minimum size=12.5pt, rotate=90},
        vnds/.style={ shape=circle, fill=black, draw, inner sep=0pt,minimum size=5pt},
        cnds/.style={ shape=rectangle, fill=white, draw, inner sep=0pt,minimum size=5pt}, 
        vndGs/.style={ shape=circle, fill=green, draw, inner sep=0pt,minimum size=5pt},
        vndGsg/.style={ shape=circle, draw=black, fill=green, draw, inner sep=0.05pt,minimum size=5pt},
        vndRs/.style={ shape=circle, fill=red, draw, inner sep=0pt,minimum size=5pt},     
        vndRsr/.style={ shape=circle, draw=red, fill=red, draw, inner sep=0pt,minimum size=5pt},     
        vndsc/.style={ shape=circle, fill=black, draw, inner sep=0pt,minimum size=5pt},
        cndsc/.style={ shape=rectangle, fill=white, draw, inner sep=0pt,minimum size=4pt}, 
        vndGsc/.style={ shape=circle, fill=green, draw=black, inner sep=0pt,minimum size=5pt},
        vndRsc/.style={ shape=circle, fill=red, draw, inner sep=0pt,minimum size=5pt}
}
    
\definecolor{mycolor1}{rgb}{0.92941,0.69412,0.12549}%
\definecolor{mycolor2}{rgb}{0.74902,0.00000,0.74902}%

\IEEEoverridecommandlockouts

\pgfplotsset{
	colormap={jetlight}{rgb = (  1.00000000,   1.00000000,   1.00000000),rgb = (  0.99607843,   0.99607843,   0.99816176),rgb = (  0.99215686,   0.99215686,   0.99644608),rgb = (  0.98823529,   0.98823529,   0.99485294),rgb = (  0.98431373,   0.98431373,   0.99338235),rgb = (  0.98039216,   0.98039216,   0.99203431),rgb = (  0.97647059,   0.97647059,   0.99080882),rgb = (  0.97254902,   0.97254902,   0.98970588),rgb = (  0.96862745,   0.96862745,   0.98872549),rgb = (  0.96470588,   0.96470588,   0.98786765),rgb = (  0.96078431,   0.96078431,   0.98713235),rgb = (  0.95686275,   0.95686275,   0.98651961),rgb = (  0.95294118,   0.95294118,   0.98602941),rgb = (  0.94901961,   0.94901961,   0.98566176),rgb = (  0.94509804,   0.94509804,   0.98541667),rgb = (  0.94117647,   0.94117647,   0.98529412),rgb = (  0.93725490,   0.93725490,   0.98529412),rgb = (  0.93333333,   0.93333333,   0.98541667),rgb = (  0.92941176,   0.92941176,   0.98566176),rgb = (  0.92549020,   0.92549020,   0.98602941),rgb = (  0.92156863,   0.92156863,   0.98651961),rgb = (  0.91764706,   0.91764706,   0.98713235),rgb = (  0.91372549,   0.91372549,   0.98786765),rgb = (  0.90980392,   0.90980392,   0.98872549),rgb = (  0.90588235,   0.90588235,   0.98970588),rgb = (  0.90196078,   0.90196078,   0.99080882),rgb = (  0.89803922,   0.89803922,   0.99203431),rgb = (  0.89411765,   0.89411765,   0.99338235),rgb = (  0.89019608,   0.89019608,   0.99485294),rgb = (  0.88627451,   0.88627451,   0.99644608),rgb = (  0.88235294,   0.88235294,   0.99816176),rgb = (  0.87843137,   0.87843137,   1.00000000),rgb = (  0.87450980,   0.87647059,   1.00000000),rgb = (  0.87058824,   0.87463235,   1.00000000),rgb = (  0.86666667,   0.87291667,   1.00000000),rgb = (  0.86274510,   0.87132353,   1.00000000),rgb = (  0.85882353,   0.86985294,   1.00000000),rgb = (  0.85490196,   0.86850490,   1.00000000),rgb = (  0.85098039,   0.86727941,   1.00000000),rgb = (  0.84705882,   0.86617647,   1.00000000),rgb = (  0.84313725,   0.86519608,   1.00000000),rgb = (  0.83921569,   0.86433824,   1.00000000),rgb = (  0.83529412,   0.86360294,   1.00000000),rgb = (  0.83137255,   0.86299020,   1.00000000),rgb = (  0.82745098,   0.86250000,   1.00000000),rgb = (  0.82352941,   0.86213235,   1.00000000),rgb = (  0.81960784,   0.86188725,   1.00000000),rgb = (  0.81568627,   0.86176471,   1.00000000),rgb = (  0.81176471,   0.86176471,   1.00000000),rgb = (  0.80784314,   0.86188725,   1.00000000),rgb = (  0.80392157,   0.86213235,   1.00000000),rgb = (  0.80000000,   0.86250000,   1.00000000),rgb = (  0.79607843,   0.86299020,   1.00000000),rgb = (  0.79215686,   0.86360294,   1.00000000),rgb = (  0.78823529,   0.86433824,   1.00000000),rgb = (  0.78431373,   0.86519608,   1.00000000),rgb = (  0.78039216,   0.86617647,   1.00000000),rgb = (  0.77647059,   0.86727941,   1.00000000),rgb = (  0.77254902,   0.86850490,   1.00000000),rgb = (  0.76862745,   0.86985294,   1.00000000),rgb = (  0.76470588,   0.87132353,   1.00000000),rgb = (  0.76078431,   0.87291667,   1.00000000),rgb = (  0.75686275,   0.87463235,   1.00000000),rgb = (  0.75294118,   0.87647059,   1.00000000),rgb = (  0.74901961,   0.87843137,   1.00000000),rgb = (  0.74509804,   0.88051471,   1.00000000),rgb = (  0.74117647,   0.88272059,   1.00000000),rgb = (  0.73725490,   0.88504902,   1.00000000),rgb = (  0.73333333,   0.88750000,   1.00000000),rgb = (  0.72941176,   0.89007353,   1.00000000),rgb = (  0.72549020,   0.89276961,   1.00000000),rgb = (  0.72156863,   0.89558824,   1.00000000),rgb = (  0.71764706,   0.89852941,   1.00000000),rgb = (  0.71372549,   0.90159314,   1.00000000),rgb = (  0.70980392,   0.90477941,   1.00000000),rgb = (  0.70588235,   0.90808824,   1.00000000),rgb = (  0.70196078,   0.91151961,   1.00000000),rgb = (  0.69803922,   0.91507353,   1.00000000),rgb = (  0.69411765,   0.91875000,   1.00000000),rgb = (  0.69019608,   0.92254902,   1.00000000),rgb = (  0.68627451,   0.92647059,   1.00000000),rgb = (  0.68235294,   0.93051471,   1.00000000),rgb = (  0.67843137,   0.93468137,   1.00000000),rgb = (  0.67450980,   0.93897059,   1.00000000),rgb = (  0.67058824,   0.94338235,   1.00000000),rgb = (  0.66666667,   0.94791667,   1.00000000),rgb = (  0.66274510,   0.95257353,   1.00000000),rgb = (  0.65882353,   0.95735294,   1.00000000),rgb = (  0.65490196,   0.96225490,   1.00000000),rgb = (  0.65098039,   0.96727941,   1.00000000),rgb = (  0.64705882,   0.97242647,   1.00000000),rgb = (  0.64313725,   0.97769608,   1.00000000),rgb = (  0.63921569,   0.98308824,   1.00000000),rgb = (  0.63529412,   0.98860294,   1.00000000),rgb = (  0.63137255,   0.99424020,   1.00000000),rgb = (  0.62745098,   1.00000000,   1.00000000),rgb = (  0.62941176,   1.00000000,   0.99411765),rgb = (  0.63149510,   1.00000000,   0.98811275),rgb = (  0.63370098,   1.00000000,   0.98198529),rgb = (  0.63602941,   1.00000000,   0.97573529),rgb = (  0.63848039,   1.00000000,   0.96936275),rgb = (  0.64105392,   1.00000000,   0.96286765),rgb = (  0.64375000,   1.00000000,   0.95625000),rgb = (  0.64656863,   1.00000000,   0.94950980),rgb = (  0.64950980,   1.00000000,   0.94264706),rgb = (  0.65257353,   1.00000000,   0.93566176),rgb = (  0.65575980,   1.00000000,   0.92855392),rgb = (  0.65906863,   1.00000000,   0.92132353),rgb = (  0.66250000,   1.00000000,   0.91397059),rgb = (  0.66605392,   1.00000000,   0.90649510),rgb = (  0.66973039,   1.00000000,   0.89889706),rgb = (  0.67352941,   1.00000000,   0.89117647),rgb = (  0.67745098,   1.00000000,   0.88333333),rgb = (  0.68149510,   1.00000000,   0.87536765),rgb = (  0.68566176,   1.00000000,   0.86727941),rgb = (  0.68995098,   1.00000000,   0.85906863),rgb = (  0.69436275,   1.00000000,   0.85073529),rgb = (  0.69889706,   1.00000000,   0.84227941),rgb = (  0.70355392,   1.00000000,   0.83370098),rgb = (  0.70833333,   1.00000000,   0.82500000),rgb = (  0.71323529,   1.00000000,   0.81617647),rgb = (  0.71825980,   1.00000000,   0.80723039),rgb = (  0.72340686,   1.00000000,   0.79816176),rgb = (  0.72867647,   1.00000000,   0.78897059),rgb = (  0.73406863,   1.00000000,   0.77965686),rgb = (  0.73958333,   1.00000000,   0.77022059),rgb = (  0.74522059,   1.00000000,   0.76066176),rgb = (  0.75098039,   1.00000000,   0.75098039),rgb = (  0.75686275,   1.00000000,   0.74117647),rgb = (  0.76286765,   1.00000000,   0.73125000),rgb = (  0.76899510,   1.00000000,   0.72120098),rgb = (  0.77524510,   1.00000000,   0.71102941),rgb = (  0.78161765,   1.00000000,   0.70073529),rgb = (  0.78811275,   1.00000000,   0.69031863),rgb = (  0.79473039,   1.00000000,   0.67977941),rgb = (  0.80147059,   1.00000000,   0.66911765),rgb = (  0.80833333,   1.00000000,   0.65833333),rgb = (  0.81531863,   1.00000000,   0.64742647),rgb = (  0.82242647,   1.00000000,   0.63639706),rgb = (  0.82965686,   1.00000000,   0.62524510),rgb = (  0.83700980,   1.00000000,   0.61397059),rgb = (  0.84448529,   1.00000000,   0.60257353),rgb = (  0.85208333,   1.00000000,   0.59105392),rgb = (  0.85980392,   1.00000000,   0.57941176),rgb = (  0.86764706,   1.00000000,   0.56764706),rgb = (  0.87561275,   1.00000000,   0.55575980),rgb = (  0.88370098,   1.00000000,   0.54375000),rgb = (  0.89191176,   1.00000000,   0.53161765),rgb = (  0.90024510,   1.00000000,   0.51936275),rgb = (  0.90870098,   1.00000000,   0.50698529),rgb = (  0.91727941,   1.00000000,   0.49448529),rgb = (  0.92598039,   1.00000000,   0.48186275),rgb = (  0.93480392,   1.00000000,   0.46911765),rgb = (  0.94375000,   1.00000000,   0.45625000),rgb = (  0.95281863,   1.00000000,   0.44325980),rgb = (  0.96200980,   1.00000000,   0.43014706),rgb = (  0.97132353,   1.00000000,   0.41691176),rgb = (  0.98075980,   1.00000000,   0.40355392),rgb = (  0.99031863,   1.00000000,   0.39007353),rgb = (  1.00000000,   1.00000000,   0.37647059),rgb = (  1.00000000,   0.99019608,   0.37254902),rgb = (  1.00000000,   0.98026961,   0.36862745),rgb = (  1.00000000,   0.97022059,   0.36470588),rgb = (  1.00000000,   0.96004902,   0.36078431),rgb = (  1.00000000,   0.94975490,   0.35686275),rgb = (  1.00000000,   0.93933824,   0.35294118),rgb = (  1.00000000,   0.92879902,   0.34901961),rgb = (  1.00000000,   0.91813725,   0.34509804),rgb = (  1.00000000,   0.90735294,   0.34117647),rgb = (  1.00000000,   0.89644608,   0.33725490),rgb = (  1.00000000,   0.88541667,   0.33333333),rgb = (  1.00000000,   0.87426471,   0.32941176),rgb = (  1.00000000,   0.86299020,   0.32549020),rgb = (  1.00000000,   0.85159314,   0.32156863),rgb = (  1.00000000,   0.84007353,   0.31764706),rgb = (  1.00000000,   0.82843137,   0.31372549),rgb = (  1.00000000,   0.81666667,   0.30980392),rgb = (  1.00000000,   0.80477941,   0.30588235),rgb = (  1.00000000,   0.79276961,   0.30196078),rgb = (  1.00000000,   0.78063725,   0.29803922),rgb = (  1.00000000,   0.76838235,   0.29411765),rgb = (  1.00000000,   0.75600490,   0.29019608),rgb = (  1.00000000,   0.74350490,   0.28627451),rgb = (  1.00000000,   0.73088235,   0.28235294),rgb = (  1.00000000,   0.71813725,   0.27843137),rgb = (  1.00000000,   0.70526961,   0.27450980),rgb = (  1.00000000,   0.69227941,   0.27058824),rgb = (  1.00000000,   0.67916667,   0.26666667),rgb = (  1.00000000,   0.66593137,   0.26274510),rgb = (  1.00000000,   0.65257353,   0.25882353),rgb = (  1.00000000,   0.63909314,   0.25490196),rgb = (  1.00000000,   0.62549020,   0.25098039),rgb = (  1.00000000,   0.61176471,   0.24705882),rgb = (  1.00000000,   0.59791667,   0.24313725),rgb = (  1.00000000,   0.58394608,   0.23921569),rgb = (  1.00000000,   0.56985294,   0.23529412),rgb = (  1.00000000,   0.55563725,   0.23137255),rgb = (  1.00000000,   0.54129902,   0.22745098),rgb = (  1.00000000,   0.52683824,   0.22352941),rgb = (  1.00000000,   0.51225490,   0.21960784),rgb = (  1.00000000,   0.49754902,   0.21568627),rgb = (  1.00000000,   0.48272059,   0.21176471),rgb = (  1.00000000,   0.46776961,   0.20784314),rgb = (  1.00000000,   0.45269608,   0.20392157),rgb = (  1.00000000,   0.43750000,   0.20000000),rgb = (  1.00000000,   0.42218137,   0.19607843),rgb = (  1.00000000,   0.40674020,   0.19215686),rgb = (  1.00000000,   0.39117647,   0.18823529),rgb = (  1.00000000,   0.37549020,   0.18431373),rgb = (  1.00000000,   0.35968137,   0.18039216),rgb = (  1.00000000,   0.34375000,   0.17647059),rgb = (  1.00000000,   0.32769608,   0.17254902),rgb = (  1.00000000,   0.31151961,   0.16862745),rgb = (  1.00000000,   0.29522059,   0.16470588),rgb = (  1.00000000,   0.27879902,   0.16078431),rgb = (  1.00000000,   0.26225490,   0.15686275),rgb = (  1.00000000,   0.24558824,   0.15294118),rgb = (  1.00000000,   0.22879902,   0.14901961),rgb = (  1.00000000,   0.21188725,   0.14509804),rgb = (  1.00000000,   0.19485294,   0.14117647),rgb = (  1.00000000,   0.17769608,   0.13725490),rgb = (  1.00000000,   0.16041667,   0.13333333),rgb = (  1.00000000,   0.14301471,   0.12941176),rgb = (  1.00000000,   0.12549020,   0.12549020),rgb = (  0.98627451,   0.12156863,   0.12156863),rgb = (  0.97242647,   0.11764706,   0.11764706),rgb = (  0.95845588,   0.11372549,   0.11372549),rgb = (  0.94436275,   0.10980392,   0.10980392),rgb = (  0.93014706,   0.10588235,   0.10588235),rgb = (  0.91580882,   0.10196078,   0.10196078),rgb = (  0.90134804,   0.09803922,   0.09803922),rgb = (  0.88676471,   0.09411765,   0.09411765),rgb = (  0.87205882,   0.09019608,   0.09019608),rgb = (  0.85723039,   0.08627451,   0.08627451),rgb = (  0.84227941,   0.08235294,   0.08235294),rgb = (  0.82720588,   0.07843137,   0.07843137),rgb = (  0.81200980,   0.07450980,   0.07450980),rgb = (  0.79669118,   0.07058824,   0.07058824),rgb = (  0.78125000,   0.06666667,   0.06666667),rgb = (  0.76568627,   0.06274510,   0.06274510),rgb = (  0.75000000,   0.05882353,   0.05882353),rgb = (  0.73419118,   0.05490196,   0.05490196),rgb = (  0.71825980,   0.05098039,   0.05098039),rgb = (  0.70220588,   0.04705882,   0.04705882),rgb = (  0.68602941,   0.04313725,   0.04313725),rgb = (  0.66973039,   0.03921569,   0.03921569),rgb = (  0.65330882,   0.03529412,   0.03529412),rgb = (  0.63676471,   0.03137255,   0.03137255),rgb = (  0.62009804,   0.02745098,   0.02745098),rgb = (  0.60330882,   0.02352941,   0.02352941),rgb = (  0.58639706,   0.01960784,   0.01960784),rgb = (  0.56936275,   0.01568627,   0.01568627),rgb = (  0.55220588,   0.01176471,   0.01176471),rgb = (  0.53492647,   0.00784314,   0.00784314),rgb = (  0.51752451,   0.00392157,   0.00392157),rgb = (  0.50000000,   0.00000000,   0.00000000)}
}
\definecolor{mittelblau}{RGB}{0, 126, 198}
\definecolor{violettblau}{cmyk}{0.9, 0.6, 0, 0}
\definecolor{rot}{RGB}{238, 28 35}
\definecolor{apfelgruen}{RGB}{140, 198, 62}
\definecolor{gelb}{RGB}{255, 229, 0}
\definecolor{orange}{RGB}{244, 111, 33}
\definecolor{pink}{RGB}{237, 0, 140}
\definecolor{lila}{RGB}{128, 10, 145}
\definecolor{hellgrau}{RGB}{224, 224, 224}
\definecolor{mittelgrau}{RGB}{128, 128, 128}
\definecolor{dunkelgrau}{RGB}{80,80,80}
\definecolor{anthrazit}{RGB}{19, 31, 31}
\definecolor{darkgreen}{RGB}{34,139,34}

\definecolor{myblue}{RGB}{80,80,160} 
\definecolor{mygreen}{RGB}{80,160,80}
\definecolor{myorgange}{RGB}{204,102,0}
\definecolor{lightblue}{RGB}{51,153,255}

\makeatletter
\newcommand{\vast}{\bBigg@{2}}

\makeatother
\begin{document}
%
%Important deadlines:
%\begin{itemize}
%	
%\item Full paper submission: December 2, 2020
% 
%\end{itemize}
%
%Other comments:
%\begin{itemize}
%\item ``$\times$'' is a bit confusing with the dimension operator? Maybe change to ``$\cdot$''?
%\item Fig.~\ref{equiSingle} legend $R_{u}$ is not defined in the paper; is it $R_{u} = R_c \cdot R_r$?
%\item Fig. legends [compress and enhance]
%\item Fig.~\ref{fcc} remove Bha sub-Fig.? or double-check the correctness of epsilon value?
%\item Fig. \ref{fig:BPFG} vs. Fig. \ref{fig:diag} bitrevorder or stage-order?
%\item Optimize Fig. positions/locations once all text is added
%\end{itemize}
%
%
%
%\newpage

	\begin{NoHyper}
		% Deep Learning-based Polar Code Construction
%		\title{Iterative Detection and Decoding of finite-length Polar Codes in Gaussian Multiple Access Channels Under Polar Belief Propagation decoder}
		\title{Iterative Detection and Decoding of Finite-Length Polar Codes in Gaussian Multiple Access Channels}

		\author{\IEEEauthorblockN{Moustafa Ebada, Sebastian Cammerer, Ahmed Elkelesh, Marvin Geiselhart and Stephan ten Brink} \thanks{This work has been supported by DFG, Germany, under grant BR 3205/5-1.}
%			\thanks{This work has been supported by DFG, Germany, under grant BR 3205/5-1.}% and BR 3205/6-1.}
			\IEEEauthorblockA{
				Institute of Telecommunications, Pfaffenwaldring 47, University of  Stuttgart, 70569 Stuttgart, Germany 
				\\\{ebada,cammerer,elkelesh,geiselhart,tenbrink\}@inue.uni-stuttgart.de
			}
		}

		%To reduce vspace between authors box and text
		\makeatletter
		\patchcmd{\@maketitle}
		{\addvspace{0.8\baselineskip}\egroup}
		{\addvspace{-0.78\baselineskip}\egroup}
		{}
		{}
		\makeatother
		%To reduce vspace between authors box and text

		\maketitle
		
		\begin{acronym}
			\acro{ECC}{error-correcting code}
			% \acro{MLD}{maximum likelihood decoding}
			\acro{HDD}{hard decision decoding}
			\acro{SDD}{soft decision decoding}
			\acro{ML}{maximum likelihood}
			\acro{GPU}{graphical processing unit}
			\acro{BP}{belief propagation}
			\acro{BPL}{belief propagation list}
			\acro{LDPC}{low-density parity-check}
			\acro{HDPC}{high density parity check}
			\acro{BER}{bit error-rate}
			\acro{SNR}{signal-to-noise-ratio}
			\acro{BPSK}{binary phase shift keying}
			\acro{AWGN}{additive white Gaussian noise}
			\acro{MSE}{mean squared error}
			\acro{LLR}{log-likelihood ratio}
			\acro{MAP}{maximum a posteriori}
			\acro{NE}{normalized error}
			\acro{BLER}{block error rate}
			\acro{PE}{processing element}
			\acro{SCL}{successive cancellation list}
			\acro{SC}{successive cancellation}
			\acro{BI-DMC}{binary input discrete memoryless channel}
			\acro{CRC}{cyclic redundancy check}
			\acro{CA-SCL}{CRC-aided successive cancellation list}
			\acro{BEC}{binary erasure channel}
			\acro{BSC}{binary symmetric channel}
			\acro{BCH}{Bose-Chaudhuri-Hocquenghem}
			\acro{RM}{Reed--Muller}
			\acro{RS}{Reed-Solomon}
			\acro{SISO}{soft-in/soft-out}
			\acro{PSCL}{partitioned successive cancellation list}
			\acro{3GPP}{$3^{\text{rd}}$ generation partnership project}
			\acro{eMBB}{enhanced mobile broadband}
			\acro{PCC}{parity-check concatenated}
			\acro{CA-polar codes}{CRC-aided polar codes}
			\acro{CN}{check nodes}
			\acro{PC}{parity-check}
			\acro{GenAlg}{Genetic Algorithm}
			\acro{AI}{artificial intelligence}
			\acro{MC}{Monte Carlo}
			\acro{CSI}{channel state information}
			\acro{PSCL}{partitioned successive cancellation list}
			\acro{5G-NR}{5G new radio}
			\acro{5G}{$5^{th}$ generation mobile communication}
			\acro{mMTC}{massive machine-type communications}
			\acro{URLLC}{ultra-reliable low-latency communications}
			\acro{SGD}{stochastic gradient descent}
			\acro{NN}{neural network}
			\acro{MI}{mutual information}		
			\acro{SPA}{sum product algorithm}
			\acro{DL}{Deep learning}
			\acro{RL}{reinforcement learning}
			\acro{MIMO}{multiple-input multiple-output}
			\acro{RCB}{random coding bound}
			\acro{GMAC}{Gaussian multiple access channel}
			\acro{MAC}{multiple access channel}
			\acro{NOMA}{non-orthogonal multiple access}
			\acro{FEC}{forward error-correction}
			\acro{IDMA}{interleave-division multiple-access}
			\acro{MUD}{multi-user detector}
			\acro{SoIC}{soft interference cancellation}
			\acro{JSC}{joint successive cancellation}
			\acro{IDD}{iterative detection and demapping}
			\acro{FCC}{frozen channel chart}
			\acro{PIC}{parallel interference cancellation}
\end{acronym}

\begin{abstract}
We consider the usage of finite-length polar codes for the \ac{GMAC} with a finite number of users. Based on the \ac{IDMA} concept, we implement an iterative detection and decoding \ac{NOMA} receiver that benefits from a low complexity, while scaling (almost) linearly with the amount of active users. 
We further show the conceptual simplicity of the \ac{BP}-based decoder in a step-by-step illustration of its construction. Beyond its conceptual simplicity, this approach benefits from an improved performance when compared to some recent work tackling the same problem, namely the setup of finite-length \ac{FEC} codes for finite-number of users. We consider the \ac{5G} polar code with a block length $N=512$ applied to both a two-user and a four-user \ac{GMAC} scenario with a sum-rate of $R_{\mathrm{sum}}=0.5$ and $R_{\mathrm{sum}}=1$, respectively. Simulation results show that a \ac{BP}-based \ac{SoIC} receiver outperforms a \ac{JSC} scheme. Finally, we investigate the effect of a concatenated repetition code which suggests that alternative polar code design rules are required in multi-user scenarios.
\end{abstract}

\section{Introduction}
Polar codes \cite{ArikanMain} are not only proven to be capacity achieving when the block length tends to infinity, they also have shown to outperform \ac{LDPC} codes in the short-length regime \cite{liva2016_2}. A key feature of polar codes is their \emph{per-bit-level} rate flexibility which can be seen as one of the reasons for their selection for the control channels of the 5G mobile communications standard \cite{polar5G2018}. 
Since their introduction, several decoding algorithms for polar codes have been developed. While \ac{SCL} decoding \cite{talvardyList} is widely used, its inherent hard-decision output is limiting its application in iterative receiver architectures, such as \ac{IDD}. Moreover, the serial nature of the algorithm puts limits to achievable latency and throughput. In contrast, iterative decoding schemes for polar codes such as \ac{BP} \cite{ArikanBP_original} and its derivative \cite{elkelesh2018belief} are based on the exchange of \emph{soft} messages and, thus, appear more promising for iterative multi-user detection.

Driven by the continuous increase of active network devices, \ac{NOMA} schemes have become a fast-growing field of research, opening up many exciting new research possibilities that require a careful reconsideration and re-evaluation of existing coding schemes \cite{dai2018survey}. For this, we focus on the application of short length communications motivated by IoT sensor networks or similar industrial applications with strict latency requirements that prevent designers from using longer codewords.
Note that, besides its conceptual simplicity in terms of orthogonalization of the channel resources, such a multi-user system may also significantly reduce the medium access latency. 
However, contrarily to the single-user case, the multi-user setup is characterized by many more code design parameters such as \ac{SNR}, flexible number of users and, potentially, non-equal transmit powers that need to be considered for the optimal code design \cite{IDMA_Wang_Cammerer}. This amount of configurable system parameters -- leading to the need of an individual code rate per user -- renders a practical deployment difficult. In particular for the short length regime, it urges the need for a multi-user coding scheme with the highest possible rate flexibility while maintaining a powerful error-rate performance at low decoding complexity.

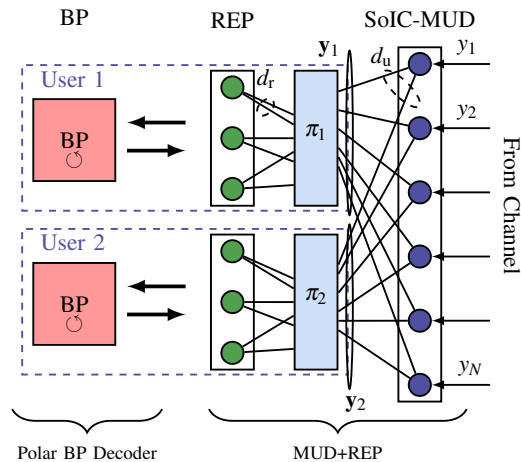
\begin{figure}[t]
	\centering
	\resizebox{!}{0.7\columnwidth}{\definecolor{mycolor22}{rgb}{0.80392,0.87843,0.96863}%

\begin{tikzpicture}[scale=0.7,xscale=-1, thick, MUDnode/.style={fill=myblue,draw,circle},   REPnode/.style={fill=mygreen,draw,circle},  VNDnode/.style={fill=myorgange,draw,circle}, CNDnode/.style={fill=myorgange,draw,rectangle}, every fit/.style={rectangle,draw,inner sep=2pt,text width=0.5cm},
>=latex]
% the vertices of MUD nodes  
\begin{scope}[start chain=going below,node distance=6mm] 
\foreach \i in {1,2,...,6}   
\node[MUDnode,on chain] (MUDn\i) [] {}; 
\end{scope}
% the vertices of Rep nodes
\begin{scope}[xshift=4cm,yshift=-0.5cm,start chain=going below,node distance=4mm] 
\foreach \i in {1,2,...,3}   
\node[REPnode,on chain] (REPn\i) [] {}; 
\end{scope}
\begin{scope}[xshift=4cm,yshift=-4cm,start chain=going below,node distance=4mm] 
\foreach \i in {4,5,...,6}   
\node[REPnode,on chain] (REPn\i) [] {}; 
\end{scope}
% the vertices of VND nodes
%\begin{scope}[xshift=5.8cm,yshift=-0.5cm,start chain=going below,node distance=4mm] 
%\foreach \i in {1,2,...,3}   
%\node[VNDnode,on chain] (VNDn\i) [] {}; 
%\end{scope}
\draw[fill=white!60!red] (8.25,-2.5) rectangle (6.5,-0.75) node[pos=.5] {BP};
%\begin{scope}[xshift=5.8cm,yshift=-4cm,start chain=going below,node distance=4mm] 
%\foreach \i in {4,5,...,6}   
%\node[VNDnode,on chain] (VNDn\i) [] {}; 
%\end{scope}
\draw[fill=white!60!red] (8.25,-6) rectangle (6.5,-4.25) node[pos=.5] {BP};
% the vertices of CND nodes
\begin{scope}[xshift=7.8cm,yshift=-10mm,start chain=going below,node distance=4mm] 
\foreach \i in {1,2}   
\node[draw=white!60!red, , fill=white!60!red, , on chain] (CNDn\i) [] {}; 
\end{scope}
\begin{scope}[xshift=7.8cm,yshift=-45mm,start chain=going below,node distance=4mm] 
\foreach \i in {3,4}   
\node[draw=white!60!red, , fill=white!60!red, , on chain] (CNDn\i) [] {}; 
\end{scope}

%\draw (5.25,-6) rectangle (3.5,-4.25) node[pos=.5] {$\circlearrowleft$};
\node [black, draw=none] at (7.375,-2.05) {$\circlearrowleft$};
\node [black, draw=none] at (7.375,-5.55) {$\circlearrowleft$};

\draw[<-, ultra thick] (6.25,-1.25) -- (5,-1.25);
\draw[->, ultra thick] (6.25,-1.85) -- (5,-1.85);

\draw[<-, ultra thick] (6.25,-4.75) -- (5,-4.75);
\draw[->, ultra thick] (6.25,-5.35) -- (5,-5.35);
%\draw (CNDn4) -- (VNDn6);
%\node[to path={(5.5, -1.5) node[midway,scale=3] {$\circlearrowleft$}  (4.5, -1.5)}]{}
% set 
\node [black,fit=(MUDn1) (MUDn6),label={[label distance=0.15cm]above: SoIC-MUD}] {};
\node [black,fit=(REPn1) (REPn3),label={[label distance=0.5cm]above: REP}] {};
\node (deinterleaver1) [fill=mycolor22, draw=black,fit=(REPn1) (REPn3),xshift=12.5mm] {$\pi_{1}$};
\node [black,fit=(REPn4) (REPn6)] {};
\node (deinterleaver2) [fill=mycolor22, draw=black,fit=(REPn4) (REPn6),xshift=12.5mm] {$\pi_{2}$};
%\node [black,fit=(VNDn1) (VNDn3)] {};
%\node [black,fit=(VNDn4) (VNDn6)] {};
%\node [black,fit=(CNDn1) (CNDn2)] {};
%\node [black,fit=(CNDn3) (CNDn4)] {};
\draw [myblue,thick,dashed] ($(deinterleaver1.north east) + (-0.2,0.1)$) rectangle ($(CNDn2.south west) + (0.5,-1)$);
\draw [myblue,thick,dashed] ($(deinterleaver2.north east) + (-0.2,0.1)$) rectangle ($(CNDn4.south west) + (0.5,-1)$);
\node [myblue, draw=none] at ($(CNDn1.north) + (-0.4,0.5)$) {User 1};
\node [myblue, draw=none] at ($(CNDn3.north) + (-0.4,0.5)$) {User 2};
% the edges 
\draw[<-] (MUDn1) -- node[xshift=0.1cm,above] {$y_1$}  +(-1.5,0); 
\draw[<-] (MUDn2) -- node[xshift=0.1cm,above] {$y_2$}  +(-1.5,0); 
\draw[<-] (MUDn3) -- +(-1.5,0); 
\draw[<-] (MUDn4) -- +(-1.5,0); 
\draw[<-] (MUDn5) -- +(-1.5,0); 
\draw[<-] (MUDn6) -- node[xshift=0.15cm,above] {$y_N$} +(-1.5,0); 

\draw (MUDn1) -- ($(deinterleaver1.east)+(0,1)$); 
\draw (MUDn1) -- ($(deinterleaver2.east)+(0,0.8)$);
\draw (MUDn2) -- ($(deinterleaver1.east)+(0,0.6)$); 
\draw (MUDn2) -- ($(deinterleaver2.east)+(0,0.6)$);
\draw (MUDn3) -- ($(deinterleaver1.east)+(0,0.2)$); 
\draw (MUDn3) -- ($(deinterleaver2.east)+(0,0.2)$);
\draw (MUDn4) -- ($(deinterleaver1.east)+(0,-0.2)$); 
\draw (MUDn4) -- ($(deinterleaver2.east)+(0,-0.2)$);
\draw (MUDn5) -- ($(deinterleaver1.east)+(0,-0.4)$); 
\draw (MUDn5) -- ($(deinterleaver2.east)+(0,-0.4)$);
\draw (MUDn6) -- ($(deinterleaver1.east)+(0,-0.6)$); 
\draw (MUDn6) -- ($(deinterleaver2.east)+(0,-0.6)$);

\draw (REPn1) -- (deinterleaver1);
\draw (REPn1) -- ($(deinterleaver1.west)+(0,0.5)$);
\draw (REPn2) -- (deinterleaver1);
\draw (REPn2) -- ($(deinterleaver1.west)+(0,-0.5)$);
\draw (REPn3) -- (deinterleaver1);
\draw (REPn3) -- ($(deinterleaver1.west)+(0,-1)$);
\draw (REPn4) -- (deinterleaver2);
\draw (REPn4) -- ($(deinterleaver2.west)+(0,0.5)$);
\draw (REPn5) -- (deinterleaver2);
\draw (REPn5) -- ($(deinterleaver2.west)+(0,-0.5)$);
\draw (REPn6) -- (deinterleaver2);
\draw (REPn6) -- ($(deinterleaver2.west)+(0,-1)$);

%\draw (REPn1) -- (VNDn1);
%\draw (REPn2) -- (VNDn2);
%\draw (REPn3) -- (VNDn3);
%\draw (REPn4) -- (VNDn4);
%\draw (REPn5) -- (VNDn5);
%\draw (REPn6) -- (VNDn6);
%
%\draw (CNDn1) -- (VNDn1);
%\draw (CNDn1) -- (VNDn2);
%\draw (CNDn1) -- (VNDn3);
%\draw (CNDn2) -- (VNDn1);
%\draw (CNDn2) -- (VNDn2);
%\draw (CNDn2) -- (VNDn3);
%\draw (CNDn3) -- (VNDn4);
%\draw (CNDn3) -- (VNDn5);
%\draw (CNDn3) -- (VNDn6);
%\draw (CNDn4) -- (VNDn4);
%\draw (CNDn4) -- (VNDn5);
%\draw (CNDn4) -- (VNDn6);

\draw [decorate,decoration={brace,amplitude=10pt,mirror,raise=4pt},yshift=0pt] ($(REPn6)+(0.5,-1)$) -- +(-5.5,0) node [black,midway,yshift=-0.8cm] {\footnotesize MUD+REP};
\draw [decorate,decoration={brace,amplitude=10pt,raise=4pt},yshift=0pt] ($(REPn6)+(1.5,-1)$) -- +(3.2,0) node [black,midway,yshift=-0.8cm] {\footnotesize Polar BP Decoder};
\node[draw=none,rotate=-90, right = 1.2cm of MUDn2] {From Channel};

\draw[rotate=-45,dashed] ($(MUDn1)+(0.65,-0.1)$) ellipse (5pt and 15pt) node[anchor=north west, xshift=-3mm,yshift=2mm,above]{$d_\mathrm{u}$};
\draw[rotate=45,dashed] ($(REPn1)+(-0.8,0.2)$) ellipse (4pt and 6pt) node[anchor=north east, xshift=0.1mm,yshift=1.5mm,above]{$d_\mathrm{r}$}; 

\draw[thick] ($(MUDn2)+(1.5,-0.1)$) ellipse (2pt and 50pt) node[anchor=north east, xshift=-3mm,yshift=10mm,above]{$\mathbf{y}_{1}$};
\draw[thick] ($(MUDn5)+(1.5,0.3)$) ellipse (2pt and 50pt) node[anchor=north east, xshift=1mm,yshift=-17mm,above]{$\mathbf{y}_{2}$};

\node [black, draw=none] at (7.375,1) {BP};
\end{tikzpicture}}
	\caption{\footnotesize Block diagram of a 2-user BP-based IDMA receiver. \label{fig:GraphModel}}\vspace{-0.7cm}
\end{figure}

Polar codes were analytically shown to achieve the channel capacity region of the two-user binary input \ac{MAC} \cite{SasogluMACpolar, OnayMACPolarSC}.
In \cite{TelatarAbbePolarMAC}, this was proven for the K-user scenario. 
However, none of these work provide an efficient decoding scheme for the case of finite code blocklength.
The authors of \cite{Marshakov} have proposed to use finite-length polar codes in an \ac{JSC}-based \ac{IDD} setup for two and four users for the special case of the unsourced MAC\footnote{Unsourced MAC is a novel	paradigm where the number of users, sharing the same codebook, can be unlimited; whereas only a finite number of users can be active on a time slot. }.

We adapt the well-known \ac{IDMA} principle \cite{IDMA_Liping} to short length polar codes with iterative decoding. \ac{IDMA} is an attractive \ac{NOMA} superposition coding scheme known to have an exceptionally good performance while maintaining a low decoding complexity using a simple \ac{MUD} based on \ac{PIC}. All users can transmit using the same channel code and modulation scheme, only a user-specific interleaver is required.
We show that polar codes under iterative decoding can be seamlessly integrated into the \emph{conventional} \ac{IDMA} scheme by replacing the \ac{LDPC} coding scheme. This promises an unseen rate flexibility on a \emph{per-bit-level} and the full compatibility to the existing 5G polar coding framework. 
In this work, we focus on (relatively) short-length \ac{NOMA} schemes that are difficult to realize with \ac{LDPC} codes \cite{IDMA_Wang_Cammerer} due to their inherent random graph structure. Because of its inherent \ac{SISO} property, \ac{BP} decoding of polar codes appears to be a promising candidate for iterative decoding and detection for multiple users allowing fully parallel decoding.

%%%%%%%%%%%%%%%%%%%%%%%%%%%%%%%%%%%%%%%%%%%%%%%%%%%%%%%%%%%%%%%%%%%%%%%%%%%%%%%%%%%%%%%%%%%%%%%%%%%%%%%%%%%%%%%%%%%%%%%%%%%%%%%%%%%%%%%%%%%%%%%%%%%%%%%%%%
\section{Preliminaries}

We provide a brief introduction to the \ac{IDMA} system model and refer the interested reader to \cite{IDMA_Wang_Cammerer} for further details about the system setup.

\subsection{IDMA System Model}

Consider the \ac{IDMA} system model with $K_\mathrm{a}$ active users, where each user encodes and decodes its data separately using a fixed polar code of code rate $R_\mathrm{c}$ followed by a repetition code of rate $R_\mathrm{r}~=~\tfrac{1}{d_\mathrm{r}}$. The total information rate is then given by $R_\mathrm{sum}~=~K_\mathrm{a}\cdot R_\mathrm{u}$, where $R_\mathrm{u}=R_\mathrm{c} \cdot R_\mathrm{r}$. After interleaving, the coded bits of the $i^{\text{th}}$ user $\mathbf{c}_i$ undergo a \ac{BPSK} mapping to $\mathbf{x}_i$. Outputs from $K_\mathrm{a}$ \emph{active} users are superimposed and transmitted over an \ac{AWGN} channel which can be expressed as
\begin{equation}
\mathbf{y} = \sum_{i=1}^{K_\mathrm{a}} \sqrt{P_i} \cdot \tilde{\mathbf{x}}_i + \mathbf{n}
\end{equation}
where $P_i$ is the received signal power of the $i^{\text{th}}$ user, $\tilde{\mathbf{x}}_i~=~\mathbf{x}_i\cdot e^{\mathrm{j}\varphi_i}$, with $\varphi_i$ being random phase shifts which are independently and uniformly distributed in $\left[ 0, \pi \right)$ and user-specific. This phase scrambling was shown to improve the performance of superimposed multi-user scenarios \cite{IDMA_Song}, especially for \ac{AWGN} channels. %{\color{red}[missing reference]}.

\subsection{Polar Codes}
\begin{figure}[t]
	\centering
	\begin{subfigure}{0.42\columnwidth}
		\resizebox{0.95\columnwidth}{!}{\tikzset{text=black, font={\fontsize{14pt}{12}\selectfont}}

\let\pgfmathMod=\pgfmathmod\relax

%\definearray{myarray}{0,1;2,3}

%\begin{document}    
\def\n{3.0}  \def\N{8.0}
\begin{tikzpicture}
\tikzset{edge/.style = {-, thick}}

\tikzset{h1/.style={preaction={%But before that
draw,yellow,-,% Draw green without any arrow head
double=yellow,
double distance=4\pgflinewidth,
}}}

\tikzset{h2/.style={preaction={%But before that
draw,green,-,% Draw green without any arrow head
double=green,
double distance=4\pgflinewidth,
}}}
  %\matrix[matrix of math nodes]{12 & \qq\\ 15 & \ww\the\mynumberone\\};
          \pgfmathtruncatemacro{\nn}{(\n + 1)}
          
\foreach \i in {1,...,\N} {
\foreach \j in {1} {
	   \ifthenelse{\i=1 \OR \i=2 \OR \i=3 \OR \i=5}
	{\node [vndRsc] (v\i\j) at (\j ,4.5-0.5*\i) {};	}
	{\node [vndsc] (v\i\j) at (\j ,4.5-0.5*\i) {};	}
   }}

    \foreach \i in {1,...,\N} {
\foreach \j in {1,...,\n} {
   \ifthenelse{\j>1}
		{\node [vndsc] (v\i\j) at (\j ,4.5-0.5*\i) {};
       		\node [cndsc] (c\i\j) at (\j+0.5 ,4.5-0.5*\i) {};}      
       		{\node [cndsc] (c\i\j) at (\j +0.5,4.5-0.5*\i) {};} 		
   }}

\foreach \i in {1,...,\N} {
\foreach \j in {4} {
	\node [vndsc] (v\i\j) at (\j ,4.5-0.5*\i) {};	
   }}

    \foreach \i in {1,...,\N} {
       \foreach \j in {1,...,\n} {
        		\draw (v\i\j)--(c\i\j);

   }}
   
      \foreach \i in {1,...,\n} {
                \pgfmathtruncatemacro{\ii}{(\i + 1)}
      \foreach \j in {1,...,\N} {
         		\draw (c\j\i)--(v\j\ii);

   }}

   \foreach \i [count=\xi] in {3,2,1} {
  \ifthenelse{\xi=1}{
  \draw (c1\i)--(v5\i);
   \draw (c2\i)--(v6\i);
   \draw (c3\i)--(v7\i);
   \draw (c4\i)--(v8\i);
   } {       \ifthenelse{\xi=2}{
     \draw (c1\i)--(v3\i);
   \draw (c2\i)--(v4\i);
   \draw (c5\i)--(v7\i);
   \draw (c6\i)--(v8\i);
   }{
     \draw (c1\i)--(v2\i);
   \draw (c3\i)--(v4\i);
   \draw (c5\i)--(v6\i);
   \draw (c7\i)--(v8\i);
     }
     }
}

%\draw [edge, black, h2] (c12)--(v32)--(c32)--(v33)--(c33)--(v43)--(c42)--(v42)--(c22)--(v23)--(c13)--(v13)--(c12);
%\draw [edge, black, h1] (c52)--(v72)--(c72)--(v73)--(c73)--(v83)--(c82)--(v82)--(c62)--(v63)--(c53)--(v53)--(c52);          
\end{tikzpicture}
%\end{document}\textsl{}
}
		\caption{\footnotesize FG; Red nodes: frozen bits }
	\end{subfigure}
	\begin{subfigure}{0.5\columnwidth}
	\vspace{-0.45cm}	\resizebox{0.95\columnwidth}{!}{\tikzset{text=black, font={\fontsize{14pt}{12}\selectfont}}

\begin{tikzpicture}
\node[] (p1) at (0.4, -2.8) {}; 
\node[] (p2) at (1.6, -1.8) {}; 
\node[] (p3) at (0.4, 2) {}; 
\node[] (p4) at (1.6, 3) {};
%\draw [gray,dashed,line width = 0.3mm] (p1)--(p2)--(p4)--(p3)--(p1); 
%\filldraw [fill=gray!20!white,draw=gray,dashed,line width = 0.3mm] (p1) rectangle (p4);
%\node[scale=1.5,gray] (p5) at (1, 3.8) {PE};
\node[] (u1) at (-2, 2) {};
\node[] (u2) at (-2, -2) {};
\node[dspnodefull,minimum size=0.2mm] (temp) at (1, -2) {};
\node[dspsquare, minimum size=0.4cm](xor) at (1,2) {};\textsl{}
\node[dspnodefull, minimum size=0.4cm](xor2) at (1,-2) {};\textsl{}
\node[] (x1) at (4, 2) {};
\node[] (x2) at (4, -2) {};
\draw[line width = 0.5mm](u1)--node[above=0.1cm] {$R_{\mathrm{in,1}}$}(xor);
\draw[line width = 0.5mm](u1)--node[below=0.2cm] {$L_{\mathrm{out,1}}$}(xor);
\draw[dspconn,line width = 0.25mm]      (-1.2,2.25) --node[above] {} (0.3,2.25);
\draw[dspconn,line width = 0.25mm]      (0.3,1.75) --node[above] {} (-1.2,1.75); 
\draw[line width = 0.5mm](temp)--(xor);
\draw[line width = 0.5mm](xor)--node[above=0.1cm] {$R_{\mathrm{out,1}}$}(x1);
\draw[line width = 0.5mm](xor)--node[below=0.2cm] {$L_{\mathrm{in,1}}$}(x1);
\node[] (shiftxy) at (2.95, 3) {};
\draw[dspconn,line width = 0.25mm]      ($(shiftxy)+(-1.2,-0.75)$) --node[above] {} ($(shiftxy)+(0.3,-0.75)$);
\draw[dspconn,line width = 0.25mm]      ($(shiftxy)+(0.3,-1.25)$) --node[above] {} ($(shiftxy)+(-1.2,-1.25)$);
\draw[line width = 0.5mm](temp)--node[above=0.1cm] {$R_{\mathrm{out,2}}$}(x2);
\draw[line width = 0.5mm](temp)--node[below=0.2cm] {$L_{\mathrm{in,2}}$}(x2);
\node[] (shiftx) at (2.95, -1) {};
\draw[dspconn,line width = 0.25mm]      ($(shiftx)+(-1.2,-0.75)$) --node[above] {} ($(shiftx)+(0.3,-0.75)$);
\draw[dspconn,line width = 0.25mm]      ($(shiftx)+(0.3,-1.25)$) --node[above] {} ($(shiftx)+(-1.2,-1.25)$);
\draw[line width = 0.5mm](u2)--node[above=0.1cm] {$R_{\mathrm{in,2}}$}(temp);
\draw[line width = 0.5mm](u2)--node[below=0.2cm] {$L_{\mathrm{out,2}}$}(temp);
\draw[dspconn,line width = 0.25mm]      (-1.2,-1.75) --node[above=0.1cm] {} (0.3,-1.75);
\draw[dspconn,line width = 0.25mm]      (0.3,-2.25) --node[above=0.1cm] {} (-1.2,-2.25);

\draw[dspconn,line width = 0.25mm]      (0.75,-1.25) --node[above=0.1cm, rotate=90] {$L_{\mathrm{in,2}}+R_{\mathrm{in,2}}$} (0.75,1.25);
\draw[dspconn,line width = 0.25mm]      (1.25,1.25) --node[above=0.1cm, rotate=-90] {$f\left(R_{\mathrm{in,1}},L_{\mathrm{in,1}}\right) $} (1.25,-1.25);

\node[]  at (-1.85,2) {};
\node[]  at (-1.85,-2) {};
\node[]  at (3.85,2) {};
\node[]  at (3.85,-2) {};
\end{tikzpicture}}
		\caption{\footnotesize Processing element}\label{fig:PE}
    \end{subfigure}
	\caption{\footnotesize Factor graph (FG) of a polar code with $N=8$ and $R_\mathrm{c}=0.5$.}
	\label{fig:BPFG}\vspace{-0.5cm}
\end{figure}
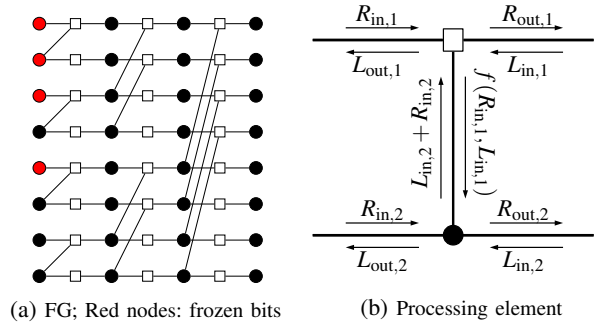
Polar codes are the first provably capacity-achieving channel codes over any arbitrary binary-input discrete memoryless channel \cite{ArikanMain}.
Channel polarization is the core idea of polar codes, where synthesized bit-channels show a polarization behavior.
The synthesized bit-channels polarize such that bit-channels become either noiseless or completely noisy and, thus, can be used for information/data transmission or frozen/known bit transmission.
A polar code of length $N$ and code dimension $k$ (i.e., $\mathcal{P}\left( k,N \right)$) can be fully described with the frozen-bits set.
Since every bit-channel can be either frozen or non-frozen, we can represent the frozen and non-frozen bits by a logical vector $\mathbb{A}$ of length $N$.
Element $i$ in the $\mathbb{A}$-vector (i.e., $a_i$) indicates whether the $i^{th}$ bit-channel is frozen or not. 
Throughout this work, $a_i=0$ means that the $i$-th bit-channel is a frozen bit, while $ a_i=1$ indicates an information bit-channel.
Thus, the polar code rate can be formulated as $R_\mathrm{c} = \frac{k}{N} = \frac{\sum_{i=1}^{N} a_i}{N}$.

Originally, the choice of frozen/non-frozen bit-channels was based on the assumption of using a \ac{SC} decoder to decode the polar code \cite{ArikanMain}.
However, for finite length polar codes, the SC decoder performs poorly in terms of error-rate.
Thus, other decoding algorithms were proposed in the literature (e.g., \ac{BP} decoder \cite{ArikanBP_original} and \ac{SCL} decoder \cite{talvardyList}) outperforming the SC decoder.
The BP decoder of polar codes is a soft-in/soft-out decoder, thus, it is very suitable for decoding concatenated coding schemes (e.g., see \cite{BP_Siegel_Concatenating,BP_sEXIT,BP_felxible}) and suitable for iterative detection and decoding. In this work, we mainly focus on BP decoding of polar codes when applied to iterative detection and decoding.

\subsection{BP Decoding and Factor Graphs}

Polar codes can be visualized as a code on a graph \cite{Forney} and, thus, iterative BP decoding (i.e., \ac{SPA}) can be used to decode polar codes \cite{ArikanBP_original}.
The Forney-style factor graph of polar codes (shown in Fig.~\ref{fig:BPFG}) consists of $\log_2 \left(N\right)$ stages. Each stage contains $\frac{N}{2}$ \acp{PE} (shown in Fig.~\ref{fig:PE}).
The channel output in terms of \ac{LLR} $ \mathbf{L_{\text{ch}} }$ is fed into the right-most stage of the factor graph.
The code structure is defined by the left-most stage of the factor graph or, in other words, the frozen and non-frozen bit-channel positions.
Then, LLR-based messages propagate over the edges of the factor graph from right-to-left ($\bf{L}$-messages) and then from left-to-right ($\bf{R}$-messages), until a maximum number of iterations $N_{\text{it,max}}$ is reached or until an early stopping condition is satisfied \cite{earlyStop}. Throughout this work, we use a $\mathbf{G}$-matrix-based early stopping condition which terminates the iterations when $\mathbf{\hat{x}}$ is equal to $\mathbf{\hat{u}} \cdot \mathbf{G}$ where $\mathbf{G}$ is the $k \times N$ polar code generator matrix.
A codeword estimate $\mathbf{\hat{x}}$ and the information bits estimate $\mathbf{\hat{u}}$ are the outputs from the BP decoder after performing a hard-decision operation on the LLR messages in the right-most and the left-most stages, respectively.

The $\bf{L}$- and $\bf{R}$-messages are updated in each \ac{PE} as follows:
\vspace*{-0.1cm}
\begin{align*}
	R_{\mathrm{out},1} &=f(R_{\mathrm{in},1},L_{\mathrm{in},2}+R_{\mathrm{in},2}) \\  R_{\mathrm{out},2} &=f(R_{\mathrm{in},1},L_{\mathrm{in},1})+R_{\mathrm{in},2}  \\    L_{\mathrm{out},1} &=f(L_{\mathrm{in},1},L_{\mathrm{in},2}+R_{\mathrm{in},2}) \\  L_{\mathrm{out},2}&=f(R_{\mathrm{in},1},L_{\mathrm{in},1})+L_{\mathrm{in},2} 
\end{align*}
where $f(L_1,L_2)=L_1 \boxplus L_2$ is commonly referred to as \emph{boxplus} operator%\cite{Hagen}
, which can be expressed as
$$ f(x,y)=x \boxplus y= \operatorname{log}\frac{1+e^{x+y}}{e^x+e^y}.$$

\section{BP-based Multi-user decoding and detection of polar codes }

Similar to \cite{IDMA_Wang_Cammerer}, the proposed decoder setup is composed of several blocks, namely:

\begin{enumerate}
\item a sub-optimal soft interference cancellation (SoIC)-based MUD block as mentioned before,
\item a repetition (REP) decoder,
\item a polar iterative decoder (i.e., BP decoder).
\end{enumerate}

\subsection{General scheme}
Assume that $\mathbf{y}$ is the received noisy channel observation of superimposed symbols $\mathbf{x}_i$.
As depicted in Fig. \ref{fig:GraphModel}, the proposed decoder setup processes $\mathbf{y}$ as follows:

\begin{enumerate}
	\item SoIC-MUD 
		\begin{itemize}
			\item Optimally: MUD would compute a posteriori probability (APP) per bit with complexity $\mathcal{O}\left(M^{K_\mathrm{a}}\right)$, where $M$ is the number of constellation symbols per user
			\item Here: a sub-optimal soft interference cancellation based low complexity MUD is implemented
			with complexity $\mathcal{O}\left(M \cdot K_\mathrm{a} \right)$ 
			\begin{itemize}
				\item $\mathbf{\hat{x}}_{i}$: feedback from single user decoder
				\item $\mathbf{y}_{j}$: forward message after SoIC $$\mathbf{y}_{j}=\mathbf{y}-\sum_{i\neq j}\mathbf{\hat{x}}_{i}$$
			\end{itemize}
		\end{itemize}
	For more details, we refer the interested reader to \cite{IDMA_Wang_Cammerer}.
	\item Demapping and De-scrambling: where log-likelihood-ratio
	(LLR) of each bit is computed by the (soft) demapper while treating the residual interference	as noise; followed by user-specific phase scrambling.
	\item Deinterleaving: according to each user's specific interleaving pattern. Accordingly, the a priori knowledge of the REP decoder comes from the extrinsic message from MUD.
	\item REP decoding: of code rate $R_\mathrm{r}=\tfrac{1}{d_\mathrm{r}}$ in case of repetition (discussed later), which enhances the signal-to-interference-plus-noise ratio (SINR) for further outer decoding. The extrinsic information of the $j^{\text{th}}$ information symbol of the
	repetition code is the summation of the LLRs of all corresponding repeated symbols.
	\item BP decoding: run a few $N_{\text{it,BP}}$ decoder-internal BP iterations (e.g., 1 or 2 iterations) to get an enhanced soft estimate of a codeword. As discussed later, more (sufficient) BP iterations are required in case of resetting the BP graph's internal memory after each BP decoding step.
	\item Repetition
	\item Interleaving
	\item[$6)$] Scrambling and remapping
	\item Repeating previous steps for $N_{\text{it,MUD}}$ times, and, thus, denoted as: $\left( N_{\text{it,BP}} \times N_{\text{it,MUD}} \right)$.
\end{enumerate}
	
Each individual \ac{BP} decoder for a single user generates an improved soft-information output, which is then used by other user detectors to further de-noise the soft input to their \ac{BP} decoders. The decoder setup is illustrated in Fig.~\ref{fig:GraphModel}.

%\footnote{We only show brief explanation and results, while we keep more detailed analysis and descriptions for an extended paper.\textcolor{red}{do we need this footnote? Sounds a bit like ``this paper is incomplete''}}

\subsection{Internal memory effects of polar BP factor graph}

\begin{figure}[t]
	\centering
	\resizebox{\columnwidth}{!}{% This file was created by matlab2tikz.
\definecolor{mycolor1}{rgb}{0.92941,0.69412,0.12549}%
\definecolor{mycolor2}{rgb}{0.74902,0.00000,0.74902}%
\begin{tikzpicture}

\begin{axis}[%
width=\columnwidth,
height=0.5\columnwidth,
at={(2.417in,1.196in)},
scale only axis,
xmin=2,
xmax=3.5,
xlabel style={font=\color{white!15!black}},
xlabel={$\frac{E_b}{N_0}$ in dB},
ymode=log,
ymin=1e-3,
ymax=2e-1,
yminorticks=true,
ylabel style={font=\color{white!15!black}},
ylabel={BLER},
axis background/.style={fill=white},
title style={font=\bfseries},
xmajorgrids,
ymajorgrids,
yminorgrids,
grid style={gray!15},
legend style={ at={(0.01,0.02)}, anchor=south west, legend cell align=left, align=left, draw=white!5!black, font=\footnotesize}
]

\addplot [color=mycolor1,  line width=2.0pt, mark=triangle, mark options={solid, mycolor1}]
table[row sep=newline, col sep=comma]{%
2.00, 9.313e-02
2.50, 2.742e-02
3.00, 5.970e-03
3.50, 1.384e-03
};
\addlegendentry{\footnotesize Clear FG before decoding}

\addplot [color=mycolor2,  line width=2.0pt, mark=square, mark options={solid, mycolor2}]
table[row sep=newline, col sep=comma]{%
2.00, 1.208e-01
2.50, 4.203e-02
3.00, 1.291e-02
3.50, 4.048e-03
};
\addlegendentry{\footnotesize Leave memory from previous codeword}

\end{axis}
\end{tikzpicture}%
}
	\caption{\footnotesize Influence of initialization of the BP factor graph for a $\mathcal{P}\left(N=512,k=256\right)$ 5G polar code; 20 BP iterations; single user scenario.}
	\label{fig:reset_fg}
\end{figure}
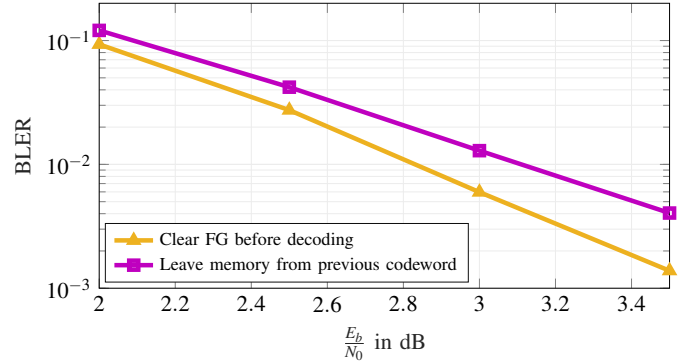

Internal LLR values of a BP factor graph instance, namely $L_{i,j}$ and $R_{i,j}$, are equivalently noted as \emph{factor graph internal states}. These states form the so-called ``internal memory'' of a graph which affects the decoding convergence to a specific codeword. To further illustrate that, two BP decoding setups are compared in Fig.~\ref{fig:reset_fg}:
\begin{enumerate}
	\item A BP decoding graph that clears the graph internal states before starting to decode a new codeword (i.e., resets the factor graph).
	\item BP decoding graph where the internal memory of it persists (i.e., internal LLR values from previous codeword are not removed while decoding a new codeword).
\end{enumerate}

As can be seen in Fig.~\ref{fig:reset_fg}, the \emph{graph internal memory initialization} plays a significant role in the convergence behavior of a specific decoding instance. This has a strong impact on the multi user turbo decoder setup.  After each MUD iteration, the BP factor graph is already at some state and, thus, the soft information is strongly affected by the previously acquired graph state. Thus, one solution might be as follows:
\begin{enumerate}
	\item run BP for a sufficient number of iterations, instead of only a few iterations
	\item pass information to MUD
	\item reset BP factor graph (resetFG), i.e., ``start over'' whenever MUD output is updated.
\end{enumerate}

The resultant new decoding setup enjoys an improved error-rate performance when compared to the previously mentioned setups, as depicted in Fig.~\ref{fig:2-user} and Fig.~\ref{fig:4-user}.

\section{Results}

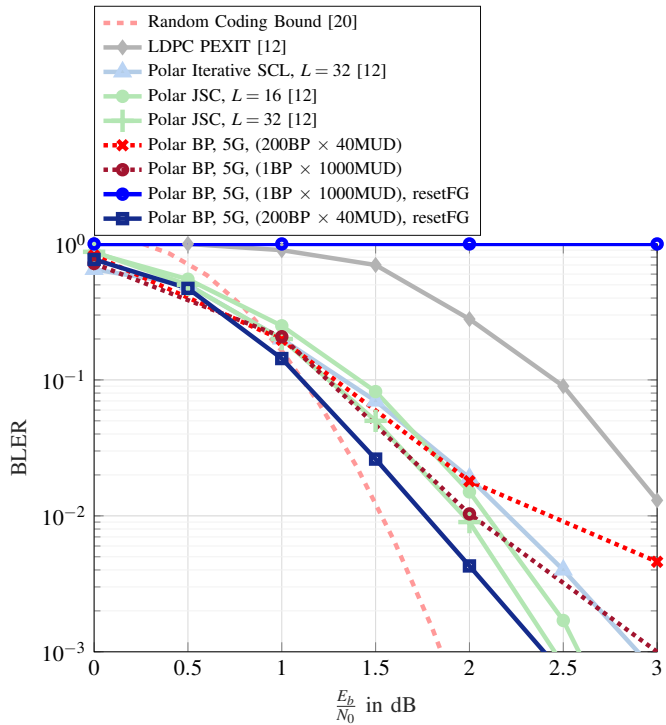
\begin{figure}[t]
	\centering
	\resizebox{\columnwidth}{!}{% This file was created by matlab2tikz.
\definecolor{mycolor1}{rgb}{0.7,0.7,0.7}%
\definecolor{mycolor2}{rgb}{0.7,0.8,0.9}%
\definecolor{mycolor3}{rgb}{0.72941,0.83137,0.95686}%
\definecolor{mycolor4}{rgb}{0.7,.9,0.7}%
\definecolor{mycolor5}{rgb}{0.46600,0.67400,0.18800}%
\definecolor{mycolor6}{rgb}{0.07843,0.16863,0.54902}%
\definecolor{mycolor7}{rgb}{0.00000,1.00000,1.00000}%
\definecolor{mycolor11}{rgb}{0.63529,0.07843,0.18431}%

%\definecolor{mycolor1}{rgb}{0.86275,0.86275,0.86275}%
%\definecolor{mycolor2}{rgb}{0.80392,0.87843,0.96863}%
%\definecolor{mycolor4}{rgb}{0.92549,0.83922,0.83922}%

\begin{tikzpicture}

\begin{axis}[%
width=\columnwidth,
height=0.725\columnwidth,
at={(2.417in,1.196in)},
scale only axis,
xmin=0,
xmax=3,
xlabel style={font=\color{white!15!black}},
xlabel={$\frac{E_b}{N_0}$ in dB},
ymode=log,
ymin=0.001,
ymax=1,
yminorticks=true,
ylabel style={font=\color{white!15!black}},
ylabel={BLER},
axis background/.style={fill=white},
xmajorgrids,
ymajorgrids,
yminorgrids,
minor grid style={draw=white!95!black},
major grid style={draw=white!85!black},
legend style={ at={(0,1.02)}, anchor=south west, legend cell align=left, align=left, draw=white!5!black}
]

\addplot [color=white!60!red, dashed, line width=2pt]
table[row sep=crcr]{%
	0.26	1.001\\
	0.4	0.85148\\
	0.6	0.59068\\
	0.8	0.33884\\
	1	0.16372\\
	1.2	0.066924\\
	1.4	0.022902\\
	1.6	0.0064961\\
	1.8	0.0015125\\
	2	0.00028637\\
	2.04	0.0001998\\
};
\addlegendentry{\footnotesize Random Coding Bound \cite{MAC_RCB_Polyanski}}

\addplot [color=mycolor1, line width=2pt, mark=diamond, mark options={solid, mycolor1}]
  table[row sep=crcr]{%
0	1\\
0.5	1\\
1	0.9\\
1.5	0.7\\
2	0.28\\
2.5	0.09\\
3	0.013\\
3.5	0.0022\\
4	0.0002\\
};
\addlegendentry{\footnotesize LDPC PEXIT \cite{Marshakov}}

\addplot [color=mycolor2, line width=2pt, mark=triangle, mark size=3pt, mark options={solid, fill=mycolor3, mycolor3}]
  table[row sep=crcr]{%
0	0.65\\
0.5	0.5\\
1	0.2\\
1.5	0.07\\
2	0.019\\
2.5	0.004\\
3	0.0007\\
3.5	0\\
4	0\\
};
\addlegendentry{\footnotesize Polar Iterative SCL, $L=32$ \cite{Marshakov}}

\addplot [color=mycolor4, line width=2pt,  mark=*, mark options={solid, fill=mycolor4, mycolor4}]
  table[row sep=crcr]{%
0	0.86\\
0.5	0.55\\
1	0.25\\
1.5	0.082\\
2	0.015\\
2.5	0.0017\\
3	7e-05\\
3.5	0\\
4	0\\
};
\addlegendentry{\footnotesize Polar JSC, $L=16$ \cite{Marshakov}}

\addplot [color=mycolor4, line width=2pt, mark size=5.0pt, mark=+, mark options={solid, mycolor4}] %[color=mycolor4, line width=2pt, mark=+, mark options={solid, fill=mycolor4, mycolor4}]
  table[row sep=crcr]{%
0	0.875\\
0.5	0.5\\
1	0.2\\
1.5	0.05\\
2	0.009\\
2.5	0.0008\\
3	0\\
3.5	0\\
4	0\\
};
\addlegendentry{\footnotesize Polar JSC, $L=32$ \cite{Marshakov}}

%\addplot [color=black, dotted, line width=2pt]
%  table[row sep=crcr]{%
%0	0.9375\\
%1	0.625\\
%2	0.260869565217391\\
%3	0.0793650793650794\\
%};
%\addlegendentry{\footnotesize Polar BP, A:5G, (200BP $\times$ 1MUD): NO MUD}

\addplot [color=red, dotted, line width=2pt, mark=x, mark options={solid}, mark size=3pt]
table[row sep=crcr]{%
	0	0.822580645161289\\
	1	0.197265625\\
	2	0.0179971489665003\\
	3	0.00459299681673488\\
};
\addlegendentry{\footnotesize Polar BP, 5G, (200BP $\times$ 40MUD)}

\addplot [color=mycolor11, dotted, line width=2pt, mark=o, mark options={solid}]
  table[row sep=crcr]{%
0	0.718309859154929\\
1	0.207818930041152\\
2	0.0103335379578473\\
3	0.000999564546138315\\
};
\addlegendentry{\footnotesize Polar BP, 5G,  (1BP $\times$ 1000MUD)}

%\addplot [color=mycolor6, dashed, line width=2pt]
%  table[row sep=crcr]{%
%0	0.879310344827585\\
%1	0.149408284023669\\
%2	0.00937790157845867\\
%3	0.00129427443743913\\
%};
%\addlegendentry{\footnotesize Polar BP, A:5G, (2BP $\times$ 1000MUD)}

\addplot [color=blue, line width=2pt, mark=o, mark options={solid, blue}]
  table[row sep=crcr]{%
0	1\\
1	1\\
2	1\\
3	1\\
}; \label{resetFG1_1000}
\addlegendentry{\footnotesize Polar BP, 5G, (1BP $\times$ 1000MUD), resetFG}

%\addplot [color=mycolor7, line width=2pt, mark=o, mark options={solid, mycolor7}]
%  table[row sep=crcr]{%
%0	0.81\\
%0.5	0.5375\\
%1	0.2\\
%1.5	0.0340677966101695\\
%2	0.00571428571428571\\
%2.5	0.000796178343949045\\
%3	0.000134625740441572\\
%};
%\addlegendentry{\footnotesize $\text{Polar BP, A:5G, (40BP $\times$ 50MUD), resetLLRs}$}

\addplot [color=mycolor6, line width=2pt, mark=square, mark options={solid, mycolor6}, mark size=2pt]
  table[row sep=crcr]{%
0	0.76923076923077\\
0.5	0.471282051282051\\
1	0.143571428571429\\
1.5	0.0261538461538462\\
2	0.00428173838578463\\
2.5	0.000687049124012367\\
3	0.00010907849397505\\
};
\addlegendentry{\footnotesize Polar BP, 5G, (200BP $\times$ 40MUD), resetFG}

\end{axis}

\end{tikzpicture}%
} 
	\caption{\footnotesize BLER performance for codelength $N=512$ and $R_\mathrm{sum}=0.5$ bits per channel use with $ K_\mathrm{a}=2 $ users over an \ac{AWGN} channel.}\label{fig:2-user}
	\vspace{-0.6cm}
\end{figure}	

Throughout this work,  all polar BP curves correspond to a polar code of length $N=512$ designed with the 5G bit-channel ordering \cite{polar5G2018},  whereas the curves from \cite{Marshakov} are optimized for their respective scheme (i.e., different polar code or $\mathbb{A}$-vector). We show results for two specific scenarios while benchmarking against the iterative \ac{SCL} and \ac{JSC} results of~\cite{Marshakov}.
\subsection{2-User GMAC}
Here, codewords from two active users are superimposed, each with a rate of $R_\mathrm{c}=\tfrac{1}{4}$ and $d_\mathrm{r}=1$ (no repetition), thus, overall information rate $R_\mathrm{sum}=\tfrac{1}{2}$. The error-rate simulation results for this setup are shown in Fig.~\ref{fig:2-user}.
		
\subsection{4-User GMAC}
Here, codewords from four active users are superimposed with $R_\mathrm{sum}=1$ fixed. 
For the case of $R_\mathrm{c}=\tfrac{1}{4}$ with $d_\mathrm{r}=1$ (no repetition) and $R_\mathrm{c}=\tfrac{1}{2}$ with $d_\mathrm{r}=2$ (two-fold repetition). 
The error-rate simulation results for this setup are shown in Fig.~\ref{fig:4-user}. 

Combined with repetition coding, the resultant scheme can be of significantly superior performance under joint polar BP-REP-SoIC decoding compared to that under JSC decoding introduced in \cite{Marshakov}. 
It is worth-mentioning that scaling up the setup for (much) more users is straightforward.

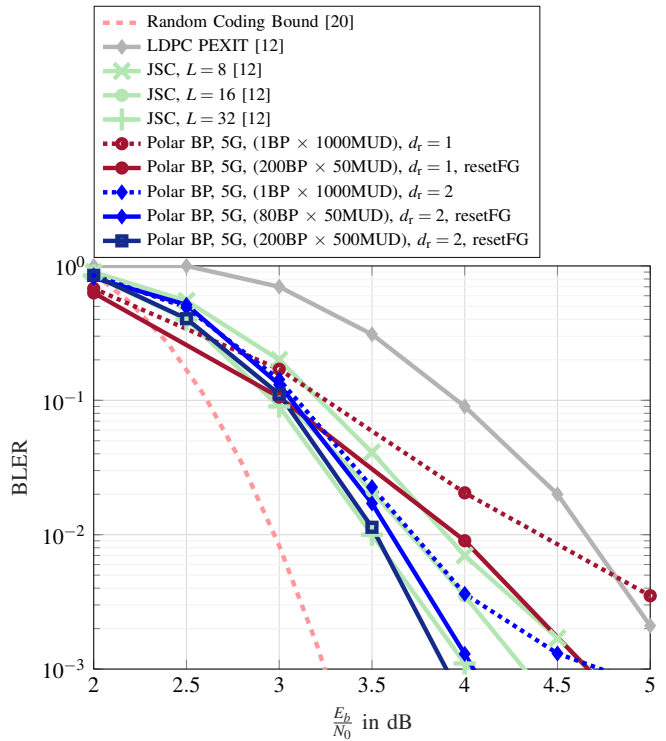
\begin{figure}[t]
	\centering
	%			\begin{subfigure}{\columnwidth}
	%				\input{tikz/4User_dr1_dr2_noResetLLRs.tikz}
	%				 	\caption{\footnotesize effect of repetition.}
	%			\end{subfigure}
	%		
	%			\begin{subfigure}{\columnwidth}
	%				\input{tikz/4User_dr1_resetLLRs_vs_noResetLLRs.tikz}
	%					\caption{\footnotesize effect of resetting LLRs.}
	%			\end{subfigure}
	%			\begin{subfigure}{\columnwidth}
	\resizebox{\columnwidth}{!}{	% This file was created by matlab2tikz.
%\definecolor{mycolor1}{rgb}{0.92549,0.83922,0.83922}%
%\definecolor{mycolor2}{rgb}{0.83137,0.81569,0.78431}%
\definecolor{mycolor3}{rgb}{0.63529,0.07843,0.18431}%
\definecolor{mycolor4}{rgb}{0.63500,0.07800,0.18400}%
\definecolor{mycolor5}{rgb}{0.00000,0.44706,0.74118}%
\definecolor{mycolor6}{rgb}{0.07843,0.16863,0.54902}%

\definecolor{mycolor1}{rgb}{0.7,.9,0.7}%
\definecolor{mycolor2}{rgb}{0.7,0.7,0.7}%
%\definecolor{mycolor1}{rgb}{0.7,0.8,0.9}%

%
\begin{tikzpicture}

\begin{axis}[%
width=\columnwidth,
height=0.725\columnwidth,
at={(2.417in,1.196in)},
scale only axis,
xmin=2,
xmax=5,
xlabel style={font=\color{white!15!black}},
xlabel={$\frac{E_b}{N_0}$ in dB},
ymode=log,
ymin=0.001,
ymax=1,
yminorticks=true,
ylabel style={font=\color{white!15!black}},
ylabel={BLER},
axis background/.style={fill=white},
title style={font=\bfseries},
title={N=512, 4 users, Rsum=1},
xmajorgrids,
ymajorgrids,
yminorgrids,
minor grid style={draw=white!95!black},
major grid style={draw=white!85!black},
legend style={ at={(0,1.02)}, anchor=south west, legend cell align=left, align=left, draw=white!5!black}
]

\addplot [color=white!60!red, dashed, line width=2pt]
table[row sep=crcr]{%
	2	0.88167\\
	2	0.88089\\
	2.2	0.53071\\
	2.4	0.26524\\
	2.6	0.10658\\
	2.8	0.033737\\
	3	0.0082876\\
	3.2	0.0015596\\
	3.4	0.0002221\\
	3.41	0.0001998\\
};
\addlegendentry{\footnotesize Random Coding Bound \cite{MAC_RCB_Polyanski}}

\addplot [color=mycolor2, line width=2pt, mark=diamond, mark options={solid, mycolor2}]
table[row sep=crcr]{%
	2	1\\
	2.5	1\\
	3	0.7\\
	3.5	0.31\\
	4	0.09\\
	4.5	0.02\\
	5	0.0021\\
};
\addlegendentry{\footnotesize LDPC PEXIT \cite{Marshakov}}

\addplot [color=mycolor1, line width=2pt, mark size=5.0pt, mark=x, mark options={solid, mycolor1}]
  table[row sep=crcr]{%
2	0.9\\
2.5	0.55\\
3	0.2\\
3.5	0.041\\
4	0.007\\
4.5	0.0017\\
5	0.00037\\
};
\addlegendentry{\footnotesize JSC, $L = 8$ \cite{Marshakov}}

\addplot [color=mycolor1, line width=2pt, mark=o, mark options={solid, mycolor1}]
  table[row sep=crcr]{%
2	0.85\\
2.5	0.4\\
3	0.11\\
3.5	0.021\\
4	0.0035\\
4.5	0.0005\\
5	8.5e-05\\
};
\addlegendentry{\footnotesize JSC, $L = 16$ \cite{Marshakov}}

\addplot [color=mycolor1, line width=2pt, mark size=5.0pt, mark=+, mark options={solid, mycolor1}]
  table[row sep=crcr]{%
2	0.85\\
2.5	0.4\\
3	0.09\\
3.5	0.01\\
4	0.0011\\
4.5	8e-05\\
5	0\\
};
\addlegendentry{\footnotesize JSC, $L = 32$ \cite{Marshakov}}

\addplot [color=mycolor3, dotted, line width=2pt, mark=o, mark options={solid, mycolor3}]
  table[row sep=crcr]{%
0	0.999999999999998\\
1	0.990196078431371\\
2	0.679054054054053\\
3	0.170302013422819\\
4	0.0204901566894335\\
5	0.00351275777700105\\
};
\addlegendentry{\footnotesize Polar BP, 5G, (1BP $\times$ 1000MUD), $d_\mathrm{r}=1$}

\addplot [color=mycolor4, line width=2pt, mark=o, mark options={solid, mycolor4}]
  table[row sep=crcr]{%
2	0.63\\
3	0.1055\\
4	0.00901785714285714\\
5	0.000330687830687831\\
};
\addlegendentry{\footnotesize Polar BP, 5G, (200BP $\times$ 50MUD), $d_\mathrm{r}=1$, resetFG}

\addplot [color=blue, dotted, line width=2pt, mark=diamond*, mark options={solid, blue}]
  table[row sep=crcr]{%
2	0.851694915254238\\
2.5	0.490196078431373\\
3	0.142857142857143\\
3.5	0.0225988700564972\\
4	0.00362976406533576\\
4.5	0.00130657468380893\\
5	0.000735704733493028\\
};
\addlegendentry{\footnotesize Polar BP, 5G, (1BP $\times$ 1000MUD), $d_\mathrm{r}=2$}

%\addplot [color=mycolor5, line width=2pt, mark=o, mark options={solid, mycolor5}]
%  table[row sep=crcr]{%
%2	0.818333333333333\\
%2.5	0.515\\
%3	0.139333333333333\\
%3.5	0.0209375\\
%4	0.00156786271450858\\
%4.5	0.000106\\
%5	1.9e-05\\
%};
%\addlegendentry{\footnotesize Polar BP, A:5G, (40BP $\times$ 50MUD), $d_\mathrm{r}=2$, resetLLRs}

\addplot [color=blue, line width=2pt, mark=diamond*, mark options={solid, blue}]
  table[row sep=crcr]{%
2	0.815\\
2.5	0.514\\
3	0.1309375\\
3.5	0.0170974576271186\\
4	0.0012966537966538\\
4.5	8.325e-05\\
5	1.8e-05\\
};
\addlegendentry{\footnotesize Polar BP, 5G, (80BP $\times$ 50MUD), $d_\mathrm{r}=2$, resetFG}

\addplot [color=mycolor6, line width=2pt, mark=square, mark options={solid, mycolor6}, mark size=2pt]
  table[row sep=crcr]{%
2	0.85\\
2.5	0.405\\
3	0.111111111111111\\
3.5	0.0113559322033898\\
4	0.000547795124623391\\
};
\addlegendentry{\footnotesize Polar BP, 5G, (200BP $\times$ 500MUD), $d_\mathrm{r}=2$, resetFG}

%\addplot [color=mycolor5, line width=2.0pt, mark=triangle, mark options={solid, mycolor5}]
%table[row sep=crcr]{%
%	2	0.823333333333333\\
%	2.5	0.565\\
%	3	0.31\\
%	3.5	0.0833333333333333\\
%	4	0.0154615384615385\\
%};
%\addlegendentry{\footnotesize $\text{Polar BP,  5G (N=1024),  (80BP X 50MUD), d}_\text{r}\text{=1}$}

\end{axis}
\end{tikzpicture}%
}
	%				 	\caption{\footnotesize best of each setup.}
	%			\end{subfigure}
	%		\vspace{-0.6cm}
	\caption{\footnotesize BLER performance for codelength $N=512$ and $R_\mathrm{sum}=1$ bits per channel use with $ K_\mathrm{a}=4 $ users over an \ac{AWGN} channel.}\label{fig:4-user}
	\vspace{-0.6cm}
\end{figure}	

\section{The Repetition Coding \emph{Dilemma}} 
It is a well-known fact that repetition coding has no coding gain for the single-user case. However, it seems as if repetition codes are very efficient for interference cancellation as pointed out in \cite{IDMA_Wang_Cammerer}. As depicted in Fig.~\ref{fig:4-user}, a joint BP-based IDD receiver can benefit from an inner repetition code of repetition factor $d_\mathrm{r}=2$ when compared to a setup that does not involve a repetition code. The resultant code/receiver setup outperforms all of the previously mentioned coding/joint-IDD setups, as can be observed in Fig.~\ref{fig:4-user}. Here, we further explore this for the case of $K_\mathrm{a}=4$ active users. %, Fig.~\ref{fig:4-user}. 

A polar code of length $N$ concatenated with an \emph{inner} repetition code of repetition factor $d_\mathrm{r}$ can be viewed as another \emph{equivalent} polar code of length $N \cdot d_\mathrm{r}$, as depicted in Fig.~\ref{fig:diag}. As can be seen, the \emph{equivalent} polar code in Fig.~\ref{N8Eq} is constructed by appending $N \cdot \left(d_\mathrm{r}-1\right)$ frozen bits to the \emph{original} polar code in Fig.~\ref{N4REP}. The equivalent polar code results in the same error-rate performance when compared to the original code setup, as depicted in Fig.~\ref{equiSingle} (i.e., compare \ref{4U512BP2REP5G} and \ref{4U1024BP1REP5eq}).

Based on that, one might deduce that the effect of concatenating an inner repetition code, with the polar code is equivalent to a  design enhancement for polar codes for the multi user setup (i.e., a better $\mathbb{A}$-vector tailored to the multi-user scenario). The performance gain with repetition is not merely a result of increasing the codelength by repetition, as can be observed from Fig. \ref{equiSingle} (i.e., compare \ref{4U1024BP1REP5G} and \ref{4U1024BP1REP5eq}). Fig.~\ref{fcc} shows a frozen channel chart (FCC) \cite{GenAlg_Journal_IEEE} for each of the \emph{equivalent} and \emph{plain} polar codes of the same length. One can easily see that the new equivalent polar code differs in many bit-positions from a polar code that follows the single user construction rules (e.g., 5G and Bhattacharyya-based polar codes). The equivalent polar code is, thus, not suitable for a traditional single user setup where it leads to a severe performance loss, as can be seen in Fig. \ref{equiSingle} (i.e., compare \ref{1U1024BP1REPeq} and \ref{1U1024BP1REP5G}).

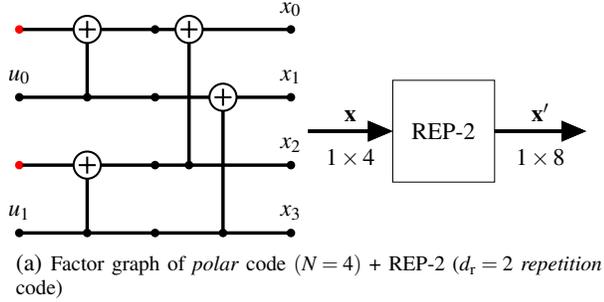
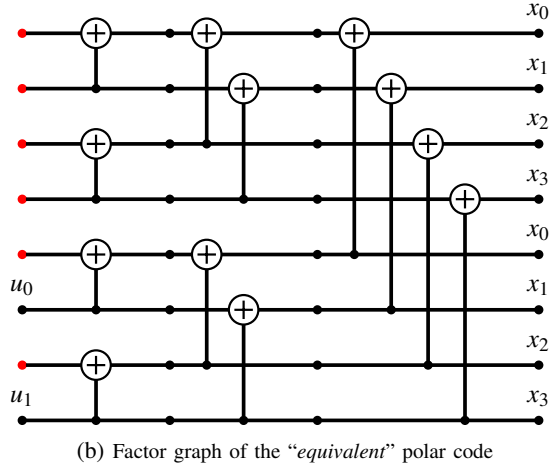
\begin{figure}[t]
	\centering\begin{subfigure}{0.88\columnwidth}
		\resizebox{0.99\columnwidth}{!}{\begin{tikzpicture}[y=1cm]
\node[dspnodefull,minimum size=1mm,red] (u1) at (0, 1) {};
\node[dspnodefull,minimum size=1mm] (u2) at (0, 0) {$u_1$};
\node[dspnodefull,minimum size=1mm] (temp) at (1, 0) {};
\node[dspadder](xor) at (1,1) {};
\node[dspnodefull,minimum size=1mm] (x1) at (2, 1) {};
\node[dspnodefull,minimum size=1mm] (x2) at (2, 0) {};
\draw[line width = 0.5mm](u1)--node[above] {}(xor);
\draw[line width = 0.5mm](temp)--(xor);
\draw[line width = 0.5mm](xor)--node[above] {}(x1);
\draw[line width = 0.5mm](temp)--node[above] {}(x2);
\draw[line width = 0.5mm](u2)--node[above] {}(temp);

\node[dspnodefull,minimum size=1mm,red] (u12) at (0, 3) {};
\node[dspnodefull,minimum size=1mm] (u22) at (0, 2) {$u_0$};
\node[dspnodefull,minimum size=1mm] (temp2) at (1, 2) {};
\node[dspadder](xor2) at (1,3) {};
\node[dspnodefull,minimum size=1mm] (x12) at (2, 3) {};
\node[dspnodefull,minimum size=1mm] (x22) at (2, 2) {};
\draw[line width = 0.5mm](u12)--node[above] {}(xor2);
\draw[line width = 0.5mm](temp2)--(xor2);
\draw[line width = 0.5mm](xor2)--node[above] {}(x12);
\draw[line width = 0.5mm](temp2)--node[above] {}(x22);
\draw[line width = 0.5mm](u22)--node[above] {}(temp2);
\node[dspnodefull,minimum size=1mm] (temp5) at (3, 0) {};
\node[dspadder](xor5) at (3,2) {};
\node[dspnodefull,minimum size=1mm] (y22) at (4, 0) {$x_3$};
\node[dspnodefull,minimum size=1mm] (y42) at (4, 2) {$x_1$};
\draw[line width = 0.5mm](x22)--node[above] {}(xor5);
\draw[line width = 0.5mm](temp5)--(xor5);
\draw[line width = 0.5mm](xor5)--node[above] {}(y42);
\draw[line width = 0.5mm](temp5)--node[above] {}(y22);
\draw[line width = 0.5mm](x2)--node[above] {}(temp5);
\node[dspnodefull,minimum size=1mm] (temp6) at (2.5, 1) {};
\node[dspadder](xor6) at (2.5,3) {};
\node[dspnodefull,minimum size=1mm] (y12) at (4, 1) {$x_2$};
\node[dspnodefull,minimum size=1mm] (y32) at (4, 3) {$x_0$};
\draw[line width = 0.5mm](x12)--node[above] {}(xor6);
\draw[line width = 0.5mm](temp6)--(xor6);
\draw[line width = 0.5mm](xor6)--node[above] {}(y32);
\draw[line width = 0.5mm](temp6)--node[above] {}(y12);
\draw[line width = 0.5mm](x1)--node[above] {}(temp6);

\draw[dspconn,line width = 0.5mm,-triangle 45](4.25, 1.5)--node[above] {$\mathbf{x}$}(5.5, 1.5);
\draw[dspconn,line width = 0.5mm,-triangle 45](4.25, 1.5)--node[below,yshift=-0.12cm] {$ 1\times 4 $}(5.5, 1.5);
\draw[fill=white] (5.5, 0.75) rectangle (7, 2.25) node[pos=.5] {REP-2};
\draw[dspconn,line width = 0.5mm,-triangle 45](7, 1.5)--node[above] {$\mathbf{x'}$}(8.35, 1.5);
\draw[dspconn,line width = 0.5mm,-triangle 45](7, 1.5)--node[below,yshift=-0.12cm] {$1\times8$}(8.35, 1.5);
%\draw[draw=black] (-0.5, -0.5) rectangle (8.5, 3.6) node[pos=.5] {};
\end{tikzpicture}}
		\caption{\footnotesize Factor graph of \emph{polar} code $\left(N=4\right)$ + REP-2 ($d_\mathrm{r}=2$   \emph{repetition} code)} \label{N4REP} \vspace{0.5cm}
	\end{subfigure}
	
	\centering\begin{subfigure}{0.8\columnwidth}
		\resizebox{\columnwidth}{!}{\begin{tikzpicture}[y=0.75cm]
\node[dspnodefull,minimum size=1mm,red] (u1) at (0, 1) {};
\node[dspnodefull,minimum size=1mm] (u2) at (0, 0) {$u_1$};
\node[dspnodefull,minimum size=1mm] (temp) at (1, 0) {};
\node[dspadder](xor) at (1,1) {};
\node[dspnodefull,minimum size=1mm] (x1) at (2, 1) {};
\node[dspnodefull,minimum size=1mm] (x2) at (2, 0) {};
\draw[line width = 0.5mm](u1)--node[above] {}(xor);
\draw[line width = 0.5mm](temp)--(xor);
\draw[line width = 0.5mm](xor)--node[above] {}(x1);
\draw[line width = 0.5mm](temp)--node[above] {}(x2);
\draw[line width = 0.5mm](u2)--node[above] {}(temp);

\node[dspnodefull,minimum size=1mm,red] (u12) at (0, 3) {};
\node[dspnodefull,minimum size=1mm] (u22) at (0, 2) {$u_0$};
\node[dspnodefull,minimum size=1mm] (temp2) at (1, 2) {};
\node[dspadder](xor2) at (1,3) {};
\node[dspnodefull,minimum size=1mm] (x12) at (2, 3) {};
\node[dspnodefull,minimum size=1mm] (x22) at (2, 2) {};
\draw[line width = 0.5mm](u12)--node[above] {}(xor2);
\draw[line width = 0.5mm](temp2)--(xor2);
\draw[line width = 0.5mm](xor2)--node[above] {}(x12);
\draw[line width = 0.5mm](temp2)--node[above] {}(x22);
\draw[line width = 0.5mm](u22)--node[above] {}(temp2);

%\node[] (u13) at (-0.5,3.5) {$\boldmath u$};
%\node[] (u13) at (7.5,3.5) {$\boldmath x$};

\node[dspnodefull,minimum size=1mm,red] (u13) at (0, 5) {};
\node[dspnodefull,minimum size=1mm,red] (u23) at (0, 4) {};
\node[dspnodefull,minimum size=1mm] (temp3) at (1, 4) {};
\node[dspadder](xor3) at (1,5) {};
\node[dspnodefull,minimum size=1mm] (x13) at (2, 5) {};
\node[dspnodefull,minimum size=1mm] (x23) at (2, 4) {};
\draw[line width = 0.5mm](u13)--node[above] {}(xor3);
\draw[line width = 0.5mm](temp3)--(xor3);
\draw[line width = 0.5mm](xor3)--node[above] {}(x13);
\draw[line width = 0.5mm](temp3)--node[above] {}(x23);
\draw[line width = 0.5mm](u23)--node[above] {}(temp3);

\node[dspnodefull,minimum size=1mm,red] (u14) at (0, 7) {};
\node[dspnodefull,minimum size=1mm,red] (u24) at (0, 6) {};
\node[dspnodefull,minimum size=1mm] (temp4) at (1, 6) {};
\node[dspadder](xor4) at (1,7) {};
\node[dspnodefull,minimum size=1mm] (x14) at (2, 7) {};
\node[dspnodefull,minimum size=1mm] (x24) at (2, 6) {};
\draw[line width = 0.5mm](u14)--node[above] {}(xor4);
\draw[line width = 0.5mm](temp4)--(xor4);
\draw[line width = 0.5mm](xor4)--node[above] {}(x14);
\draw[line width = 0.5mm](temp4)--node[above] {}(x24);
\draw[line width = 0.5mm](u24)--node[above] {}(temp4);

\node[dspnodefull,minimum size=1mm] (temp5) at (3, 0) {};
\node[dspadder](xor5) at (3,2) {};
\node[dspnodefull,minimum size=1mm] (y22) at (4, 0) {};
\node[dspnodefull,minimum size=1mm] (y42) at (4, 2) {};
\draw[line width = 0.5mm](x22)--node[above] {}(xor5);
\draw[line width = 0.5mm](temp5)--(xor5);
\draw[line width = 0.5mm](xor5)--node[above] {}(y42);
\draw[line width = 0.5mm](temp5)--node[above] {}(y22);
\draw[line width = 0.5mm](x2)--node[above] {}(temp5);

\node[dspnodefull,minimum size=1mm] (temp6) at (2.5, 1) {};
\node[dspadder](xor6) at (2.5,3) {};
\node[dspnodefull,minimum size=1mm] (y12) at (4, 1) {};
\node[dspnodefull,minimum size=1mm] (y32) at (4, 3) {};
\draw[line width = 0.5mm](x12)--node[above] {}(xor6);
\draw[line width = 0.5mm](temp6)--(xor6);
\draw[line width = 0.5mm](xor6)--node[above] {}(y32);
\draw[line width = 0.5mm](temp6)--node[above] {}(y12);
\draw[line width = 0.5mm](x1)--node[above] {}(temp6);

\node[dspnodefull,minimum size=1mm] (temp7) at (3, 4) {};
\node[dspadder](xor7) at (3,6) {};
\node[dspnodefull,minimum size=1mm] (y6) at (4, 4) {};
\node[dspnodefull,minimum size=1mm] (y8) at (4, 6) {};
\draw[line width = 0.5mm](x24)--node[above] {}(xor7);
\draw[line width = 0.5mm](temp7)--(xor7);
\draw[line width = 0.5mm](xor7)--node[above] {}(y8);
\draw[line width = 0.5mm](temp7)--node[above] {}(y6);
\draw[line width = 0.5mm](x23)--node[above] {}(temp7);

\node[dspnodefull,minimum size=1mm] (temp8) at (2.5, 5) {};
\node[dspadder](xor8) at (2.5,7) {};
\node[dspnodefull,minimum size=1mm] (y5) at (4, 5) {};
\node[dspnodefull,minimum size=1mm] (y7) at (4, 7) {};
\draw[line width = 0.5mm](x14)--node[above] {}(xor8);
\draw[line width = 0.5mm](temp8)--(xor8);
\draw[line width = 0.5mm](xor8)--node[above] {}(y7);
\draw[line width = 0.5mm](temp8)--node[above] {}(y5);
\draw[line width = 0.5mm](x13)--node[above] {}(temp8);

\node[dspnodefull,minimum size=1mm] (temp9) at (4.5, 3) {};
\node[dspadder](xor9) at (4.5,7) {};
\node[dspnodefull,minimum size=1mm] (z1) at (7, 3) {$x_0$};
\node[dspnodefull,minimum size=1mm] (z5) at (7, 7) {$x_0$};
\draw[line width = 0.5mm](y7)--node[above] {}(xor9);
\draw[line width = 0.5mm](temp9)--(xor9);
\draw[line width = 0.5mm](xor9)--node[above] {}(z5);
\draw[line width = 0.5mm](temp9)--node[above] {}(z1);
\draw[line width = 0.5mm](y32)--node[above] {}(temp9);

\node[dspnodefull,minimum size=1mm] (temp10) at (5, 2) {};
\node[dspadder](xor10) at (5,6) {};
\node[dspnodefull,minimum size=1mm] (z2) at (7, 2) {$x_1$};
\node[dspnodefull,minimum size=1mm] (z6) at (7, 6) {$x_1$};
\draw[line width = 0.5mm](y8)--node[above] {}(xor10);
\draw[line width = 0.5mm](temp10)--(xor10);
\draw[line width = 0.5mm](xor10)--node[above] {}(z6);
\draw[line width = 0.5mm](temp10)--node[above] {}(z2);
\draw[line width = 0.5mm](y42)--node[above] {}(temp10);

\node[dspnodefull,minimum size=1mm] (temp11) at (5.5, 1) {};
\node[dspadder](xor11) at (5.5,5) {};
\node[dspnodefull,minimum size=1mm] (z3) at (7, 1) {$x_2$};
\node[dspnodefull,minimum size=1mm] (z7) at (7, 5) {$x_2$};
\draw[line width = 0.5mm](y5)--node[above] {}(xor11);
\draw[line width = 0.5mm](temp11)--(xor11);
\draw[line width = 0.5mm](xor11)--node[above] {}(z7);
\draw[line width = 0.5mm](temp11)--node[above] {}(z3);
\draw[line width = 0.5mm](y12)--node[above] {}(temp11);

\node[dspnodefull,minimum size=1mm] (temp12) at (6, 0) {};
\node[dspadder](xor12) at (6,4) {};
\node[dspnodefull,minimum size=1mm] (z4) at (7, 0) {$x_3$};
\node[dspnodefull,minimum size=1mm] (z8) at (7, 4) {$x_3$};
\draw[line width = 0.5mm](y6)--node[above] {}(xor12);
\draw[line width = 0.5mm](temp12)--(xor12);
\draw[line width = 0.5mm](xor12)--node[above] {}(z8);
\draw[line width = 0.5mm](temp12)--node[above] {}(z4);
\draw[line width = 0.5mm](y22)--node[above] {}(temp12);

%\node[] (ch1) at (8.5, 3.5) {};
%\node[] (ch2) at (7.2, 3.5) {};
%%\draw[dspconn,line width = 0.5mm,-triangle 45](ch1)--node[above] {$L_{\mathrm{ch}}$}(ch2);
	
%\node[] (src1) at (-1.7, 3.5) {};
%\node[] (src2) at (-0.2, 3.5) {};
%%\draw[dspconn,line width = 0.5mm,-triangle 45](src1)--node[above] {$L_{\mathrm{frozen}}$}(src2);
%
%\node[] (Lmsg1) at (7, 8) {};
%\node[] (Lmsg2) at (0, 8) {};
%%\draw[dspconn,line width = 0.5mm,-triangle 45](Lmsg1)--node[above] {$\mathbf{L}$-messages propagation}(Lmsg2);
%
%\node[] (Rmsg1) at (0, -1) {};
%\node[] (Rmsg2) at (7, -1) {};
%\draw[dspconn,line width = 0.5mm,-triangle 45](Rmsg1)--node[below] {$\mathbf{R}$-messages propagation}(Rmsg2);
%\draw[draw=black] (-0.5, -0.5) rectangle (7.5, 8) node[pos=.5] {};
\end{tikzpicture}}
		\caption{\footnotesize Factor graph of the ``\emph{equivalent}'' polar code} \label{N8Eq}
	\end{subfigure}
	\caption{\footnotesize An equivalent perspective of the polar code concatenated with an inner repetition code.} \label{fig:diag}
\end{figure}

\begin{figure}[H]
	\centering
	\resizebox{\columnwidth}{!}{% This file was created by matlab2tikz.
\begin{tikzpicture}

\begin{axis}[%
width=\columnwidth,
height=0.25\columnwidth,
at={(1.267in,3.103in)},
scale only axis,
axis on top,
xmin=0.5,
xmax=64.5,
xtick={\empty},
y dir=reverse,
ymin=0.5,
ymax=16.5,
ytick={ 1, 16},
axis background/.style={fill=white},
title style={font=\bfseries},
title={Polar code designed with Bhattacharrya for a BEC$\left( \epsilon=0.3172 \right)$ $\left(N=1024\right)$},
legend style={legend cell align=left, align=left, draw=white!15!black}
]
\addplot [forget plot] graphics [xmin=0.5, xmax=64.5, ymin=0.5, ymax=16.5] {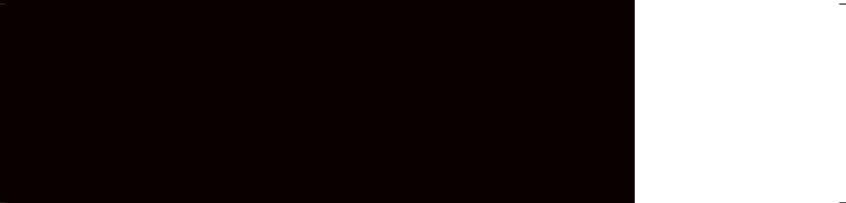}; \vspace{2.25cm}
\end{axis}

\begin{axis}[%
width=\columnwidth,
height=0.25\columnwidth,
at={(1.267in,1.792in)},
scale only axis,
axis on top,
xmin=0.5,
xmax=64.5,
xtick={\empty},
y dir=reverse,
ymin=0.5,
ymax=16.5,
ytick={ 1, 16},
axis background/.style={fill=white},
title style={font=\bfseries},
title={Polar code designed according to 5G specification $\left(N=1024\right)$},
legend style={legend cell align=left, align=left, draw=white!15!black}
]
\addplot [forget plot] graphics [xmin=0.5, xmax=64.5, ymin=0.5, ymax=16.5] {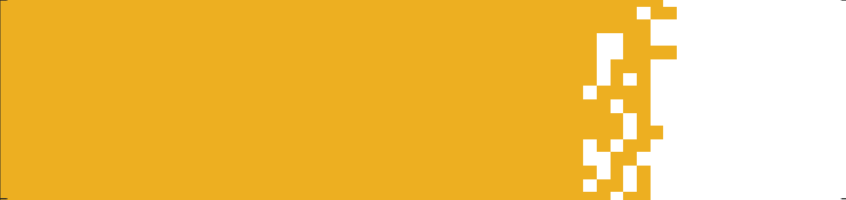};
\end{axis}

\begin{axis}[%
width=\columnwidth,
height=0.25\columnwidth,
at={(1.267in,0.481in)},
scale only axis,
axis on top,
xmin=0.5,
xmax=64.5,
xtick={\empty},
y dir=reverse,
ymin=0.5,
ymax=16.5,
ytick={ 1, 16},
axis background/.style={fill=white},
title style={font=\bfseries},
title={Polar code $\left(N=1024\right)$ \emph{equivalent} to polar code $\left(N=512\right)$ with 2-repetition},
legend style={legend cell align=left, align=left, draw=white!15!black}
]
\addplot [forget plot] graphics [xmin=0.5, xmax=64.5, ymin=0.5, ymax=16.5] {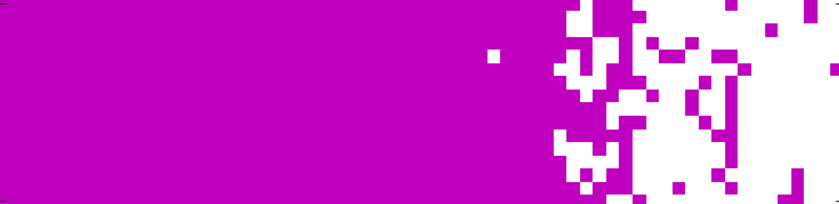};
\end{axis}
\end{tikzpicture}%
}
	\caption{\footnotesize Difference between the equivalent $\mathbb{A}$-vector and a good single-user code design (i.e., 5G-based $\mathbb{A}$-vector) visualized using a frozen channel chart (FCC) \cite{GenAlg_Journal_IEEE}. The $1024$ bit positions are plotted over a $16\times64$ matrix. Note that the bit-channels are sorted with decreasing Bhattacharyya parameter value. White: non-frozen; colored: frozen.} \label{fcc}
\end{figure}

\section{Conclusion}

In this paper, we proposed the usage of \ac{BP} decoders for iterative detection and decoding of polar codes in a multi-user scenario. Combined with repetition coding, the resulting scheme can compete with, and even outperform, \ac{JSC} based \ac{IDD} \cite{Marshakov} for two and four users, while the inherent parallel nature of polar \ac{BP} decoding may lead to more desirable hardware implementations. Further analysis is required to investigate the case for a massive number of users. 

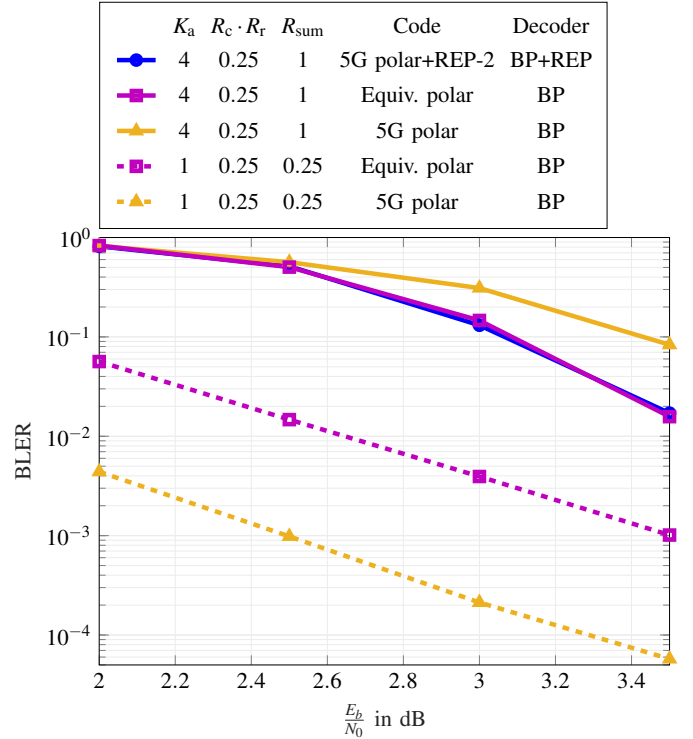
\begin{figure}[t]
	\centering
	\resizebox{\columnwidth}{!}{% This file was created by matlab2tikz.
\definecolor{mycolor1}{rgb}{0.92941,0.69412,0.12549}%
\definecolor{mycolor2}{rgb}{0.74902,0.00000,0.74902}%
\begin{tikzpicture}

\begin{axis}[%
width=\columnwidth,
height=0.75\columnwidth,
at={(2.417in,1.196in)},
scale only axis,
xmin=2,
xmax=3.5,
xlabel style={font=\color{white!15!black}},
xlabel={$\frac{E_b}{N_0}$ in dB},
ymode=log,
ymin=0.00005,
ymax=1,
yminorticks=true,
ylabel style={font=\color{white!15!black}},
ylabel={BLER},
axis background/.style={fill=white},
title style={font=\bfseries},
xmajorgrids,
ymajorgrids,
yminorgrids,
grid style={gray!15},
legend style={ at={(0,1.02)}, anchor=south west, legend cell align=left, align=left, draw=white!5!black, font=\footnotesize}
]

\addplot [color=blue, line width=2pt, mark=o, mark options={solid, blue}]
table[row sep=crcr]{%
	2	0.815\\
	2.5	0.514\\
	3	0.1309375\\
	3.5	0.0170974576271186\\
	4	0.0012966537966538\\
	4.5	8.325e-05\\
	5	1.8e-05\\
};
\label{4U512BP2REP5G}
%\addlegendentry{\footnotesize $\text{4-Users + REP, N=512, A:5G, (80BP X 50MUD), d}_\text{r}\text{=2}$}

\addplot [color=mycolor1,  line width=2.0pt, mark=triangle, mark options={solid, mycolor1}]
table[row sep=crcr]{%
	2	0.823333333333333\\
	2.5	0.565\\
	3	0.31\\
	3.5	0.0833333333333333\\
	4	0.0154615384615385\\
};
\label{4U1024BP1REP5G}
%\addlegendentry{\footnotesize $\text{4-Users, N=1024, A:5G,  (80BP X 50MUD), d}_\text{r}\text{=1}$}

\addplot [color=mycolor2,  line width=2.0pt, mark=square, mark options={solid, mycolor2}]
table[row sep=crcr]{%
	2	0.83\\
	2.5	0.50375\\
	3	0.146428571428571\\
	3.5	0.015748031496063\\
	4	0.001355\\
};
\label{4U1024BP1REP5eq}
%\addlegendentry{\footnotesize $\text{4-Users, N=1024, A:equivalent to Polar512+REP,  (80BP X 50MUD), d}_\text{r}\text{=1}$}

\addplot [color=mycolor1, dashed,line width=2pt, mark=triangle, mark options={solid, mycolor1}]
table[row sep=crcr]{%
	2	0.0043941\\
	2.5	0.0009827\\
	3	0.0002126\\
	3.5	5.74e-05\\
	4	3.53e-05\\
};
\label{1U1024BP1REP5G}
%\addlegendentry{\footnotesize 1-User, (256, 1024)- polar code, 5G}

\addplot [color=mycolor2, dashed,line width=2pt, mark size=2pt, mark=square, mark options={solid, mycolor2}]
table[row sep=crcr]{%
	2	0.0564814814814815\\
	2.5	0.0147058823529412\\
	3	0.00393700787401575\\
	3.5	0.00101385603244339\\
	4	0\\
};
\label{1U1024BP1REPeq}
%\addlegendentry{\footnotesize 1-User,  (256, 1024)- polar code, equivalent to (256, 512)- polar + REP-2}
\coordinate (legend) at (axis description cs:0,1.02);

\end{axis}

\matrix [
draw,
matrix of nodes,
anchor=south west,
] at (legend) {
	                      & $K_\mathrm{a}$   & $R_\mathrm{c}\cdot R_\mathrm{r}$ & $R_\mathrm{sum}$ & Code           & Decoder\\
	\ref{4U512BP2REP5G}   & 4        & $0.25$  & $1$       & 5G polar+REP-2 & BP+REP    \\
	\ref{4U1024BP1REP5eq} & 4        & $0.25$  & $1$       & Equiv. polar   & BP    \\
	\ref{4U1024BP1REP5G}  & 4        & $0.25$  & $1$       & 5G polar       & BP   \\
	\ref{1U1024BP1REPeq}  & 1        & $0.25$  & $0.25$    & Equiv. polar   & BP  \\ 
	\ref{1U1024BP1REP5G}  & 1        & $0.25$  & $0.25$    & 5G polar       & BP   \\
};

\end{tikzpicture}%
}
	\caption{\footnotesize 
	Error-rate of a conventional polar code concatenated with an inner repetition code compared to that of the \emph{equivalent} longer polar code, similar to Fig.~\ref{fig:diag}. Single-user performance is also plotted to show difference to multi-user setup (i.e., in a multi-user setup: equivalent code is superior in performance to that of a typically constructed polar code without repetition; whereas for a single-user setup it is exactly the opposite). Sum-rates of both setups are obviously not comparable. All curves shown for codelength $N=1024$.
	}
	\label{equiSingle}
\end{figure}

We also shed some light on the gains by the usage of a concatenated repetition code and verified its effectiveness in Monte-Carlo simulations, a theoretical explanation remains, however, an open problem. We concluded that this repetition code corresponds to a different selection of frozen bit-positions (different $\mathbb{A}$-vector) in the polar code design, suggesting that the multi-user detection can benefit from a different polar code design than the single-user case. Finding such an optimized polar code construction is an interesting open future work and an important step towards efficient deployment of multi-user polar coding with iterative detection and decoding.

%\section{Insights and Future Avenues} 
%\ac{SISO}-\ac{BP} decoding of polar codes proves to be an efficient multi-user detection method. Combined with repetition coding, the resultant scheme can be of significantly superior performance under joint polar BP-REP-SoIC decoding scheme {\color{red} compared to what? JSC of \cite{Marshakov}?}. 
%Further analysis is required to investigate the case for a massive number of users.

\bibliographystyle{IEEEtran}
\bibliography{references}

\begin{thebibliography}{10}
\providecommand{\url}[1]{#1}
\csname url@samestyle\endcsname
\providecommand{\newblock}{\relax}
\providecommand{\bibinfo}[2]{#2}
\providecommand{\BIBentrySTDinterwordspacing}{\spaceskip=0pt\relax}
\providecommand{\BIBentryALTinterwordstretchfactor}{4}
\providecommand{\BIBentryALTinterwordspacing}{\spaceskip=\fontdimen2\font plus
\BIBentryALTinterwordstretchfactor\fontdimen3\font minus
  \fontdimen4\font\relax}
\providecommand{\BIBforeignlanguage}[2]{{%
\expandafter\ifx\csname l@#1\endcsname\relax
\typeout{** WARNING: IEEEtran.bst: No hyphenation pattern has been}%
\typeout{** loaded for the language `#1'. Using the pattern for}%
\typeout{** the default language instead.}%
\else
\language=\csname l@#1\endcsname
\fi
#2}}
\providecommand{\BIBdecl}{\relax}
\BIBdecl

\bibitem{ArikanMain}
E.~{Arıkan}, ``{Channel Polarization: A Method for Constructing
  Capacity-Achieving Codes for Symmetric Binary-Input Memoryless Channels},''
  \emph{IEEE Trans. Inf. Theory}, vol.~55, no.~7, pp. 3051--3073, Jul. 2009.

\bibitem{liva2016_2}
M.~C. {Coşkun}, G.~Durisi, T.~Jerkovits, G.~Liva, W.~E. Ryan, B.~Stein, and
  F.~Steiner, ``{Efficient Error-Correcting Codes in the Short Blocklength
  Regime},'' \emph{Physical Communication}, vol.~34, pp. 66--79, Jun. 2019.

\bibitem{polar5G2018}
\BIBentryALTinterwordspacing
``{Technical Specification Group Radio Access Network},'' \emph{3GPP, 2018, TS
  38.212 V.15.1.1.} [Online]. Available:
  \url{http://www.3gpp.org/ftp/Specs/archive/38_series/38.212/}
\BIBentrySTDinterwordspacing

\bibitem{talvardyList}
I.~Tal and A.~Vardy, ``{List Decoding of Polar Codes},'' \emph{IEEE Trans. Inf.
  Theory}, vol.~61, no.~5, pp. 2213--2226, May 2015.

\bibitem{ArikanBP_original}
E.~Arıkan, ``{Polar Codes: A Pipelined Implementation},'' \emph{Proc. 4th
  ISBC}, pp. 11--14, 2010.

\bibitem{elkelesh2018belief}
A.~Elkelesh, M.~Ebada, S.~Cammerer, and S.~ten Brink, ``{Belief Propagation
  List Decoding of Polar Codes},'' \emph{IEEE Commun. Lett.}, vol.~22, no.~8,
  pp. 1536--1539, Aug. 2018.

\bibitem{dai2018survey}
L.~Dai, B.~Wang, Z.~Ding, Z.~Wang, S.~Chen, and L.~Hanzo, ``{A Survey of
  Non-Orthogonal Multiple Access for 5G},'' \emph{IEEE Commun. Surveys \&
  Tutorials}, vol.~20, no.~3, pp. 2294--2323, 2018.

\bibitem{IDMA_Wang_Cammerer}
X.~{Wang}, S.~{Cammerer}, and S.~{ten Brink}, ``{Near-Capacity Detection and
  Decoding: Code Design for Dynamic User Loads in Gaussian Multiple Access
  Channels},'' \emph{IEEE Trans. Commun.}, vol.~67, no.~11, pp. 7417--7430,
  2019.

\bibitem{SasogluMACpolar}
E.~{Şaşoğlu}, E.~{Telatar}, and E.~M. {Yeh}, ``{Polar Codes for the Two-User
  Multiple-Access Channel},'' \emph{IEEE Trans. Inf. Theory}, vol.~59, no.~10,
  pp. 6583--6592, 2013.

\bibitem{OnayMACPolarSC}
S.~{Önay}, ``{Successive Cancellation Decoding of Polar Codes for the Two-User
  Binary-Input MAC},'' in \emph{IEEE Inter. Symp. Inf. Theory (ISIT)}, 2013,
  pp. 1122--1126.

\bibitem{TelatarAbbePolarMAC}
E.~{Abbe} and E.~{Telatar}, ``{Polar Codes for the $m$-User Multiple Access
  Channel},'' \emph{IEEE Trans. Inf. Theory}, vol.~58, no.~8, pp. 5437--5448,
  2012.

\bibitem{Marshakov}
E.~{Marshakov}, G.~{Balitskiy}, K.~{Andreev}, and A.~{Frolov}, ``{A Polar Code
  Based Unsourced Random Access for the Gaussian MAC},'' in \emph{IEEE 90th
  Vehicular Technology Conference (VTC2019-Fall)}, 2019, pp. 1--5.

\bibitem{IDMA_Liping}
{Li Ping}, {Lihai Liu}, {Keying Wu}, and W.~K. {Leung}, ``{Interleave Division
  Multiple-Access},'' \emph{IEEE Trans. on Wireless Commun.}, vol.~5, no.~4,
  pp. 938--947, 2006.

\bibitem{IDMA_Song}
G.~{Song} and J.~{Cheng}, ``{Distance Enumerator Analysis for
  Interleave-Division Multi-User Codes},'' \emph{IEEE Trans. Inf. Theory},
  vol.~62, no.~7, pp. 4039--4053, 2016.

\bibitem{BP_Siegel_Concatenating}
J.~Guo, M.~Qin, A.~G. i~Fàbregas, and P.~H. Siegel, ``{Enhanced Belief
  Propagation Decoding of Polar Codes through Concatenation},'' in \emph{IEEE
  Inter. Symp. Inf. Theory (ISIT)}, Jun. 2014, pp. 2987--2991.

\bibitem{BP_sEXIT}
A.~Elkelesh, M.~Ebada, S.~Cammerer, and S.~ten Brink, ``{Improving Belief
  Propagation Decoding of Polar Codes Using Scattered EXIT Charts},'' in
  \emph{IEEE Inf. Theory Workshop (ITW)}, Sep. 2016, pp. 91--95.

\bibitem{BP_felxible}
------, ``{Flexible Length Polar Codes through Graph Based Augmentation},'' in
  \emph{IEEE Inter. ITG Conf. on Syst., Commun. and Coding (SCC)}, Feb. 2017.

\bibitem{Forney}
G.~D. Forney, ``{Codes on Graphs: Normal Realizations},'' \emph{IEEE Trans.
  Inf. Theory}, vol.~47, no.~2, pp. 520--548, Feb. 2001.

\bibitem{earlyStop}
B.~Yuan and K.~K. Parhi, ``{Early Stopping Criteria for Energy-Efficient
  Low-Latency Belief-Propagation Polar Code Decoders},'' \emph{IEEE Trans. Sig.
  Process.}, vol.~62, no.~24, pp. 6496--6506, Dec. 2014.

\bibitem{MAC_RCB_Polyanski}
Y.~{Polyanskiy}, ``{A Perspective on Massive Random-Access},'' in \emph{IEEE
  Inter. Symp. Inf. Theory (ISIT)}, 2017, pp. 2523--2527.

\bibitem{GenAlg_Journal_IEEE}
A.~{Elkelesh}, M.~{Ebada}, S.~{Cammerer}, and S.~{ten Brink},
  ``{Decoder-Tailored Polar Code Design Using the Genetic Algorithm},''
  \emph{IEEE Trans. Commun.}, vol.~67, no.~7, pp. 4521--4534, Jul. 2019.

\end{thebibliography}

\end{NoHyper}
\end{document}